\documentclass[10pt,journal,letterpaper,twocolumn,twoside]{IEEEtran} 

\usepackage{amsmath,amsfonts,bm,amssymb,psfrag,ifthen,color,subfigure}
\usepackage{mathrsfs}

\usepackage[justification=centering]{caption}

\usepackage[dvips]{epsfig}
\usepackage[dvips]{graphicx}
\usepackage{amsmath,amsfonts,bm,amssymb,psfrag,ifthen,color,subfigure}
\usepackage{algpseudocode}
\usepackage{algorithm}
\usepackage{algorithmicx}
\usepackage{setspace}
\usepackage{cite}
\usepackage{url}
\usepackage{longtable}
\usepackage{stfloats}  
\usepackage{graphics,booktabs,color,subfigure}
\usepackage{float}
\usepackage{diagbox}

\begin{document}%

\title{   Robust  Beamforming Design  for Covert Communications}
 \author{Shuai Ma, \emph{Member, IEEE}, Yunqi Zhang, Hang Li, Songtao Lu, \emph{Member, IEEE}, Naofal Al-Dhahir, \emph{Fellow, IEEE}, Sha Zhang and Shiyin Li.

 \thanks{S. Ma is with the School of Information and Control Engineering, China
University of Mining and Technology, Xuzhou 221116, China, and also with
National Mobile Communications Research Laboratory, Southeast University, Nanjing 210096, China
(e-mail: mashuai001@cumt.edu.cn).}

\thanks{Y. Zhang is with the School of Information and Control   Engineering, China
University of Mining and Technology, Xuzhou 221116,
China (e-mail: ts19060151p31@cumt.edu.cn).}
\thanks{H. Li is with the Shenzhen Research Institute of Big Data, Shenzhen 518172, Guangdong, China. (email: hangdavidli@163.com).}
\thanks{S. Lu is with the IBM Thomas J. Watson Research Center, Yorktown
Heights, NY 10598 USA (e-mail: songtao@ibm.com).}
\thanks{N. Al-Dhahir is with the Electrical and Computer Engineering Department, University of
Texas at Dallas, Dallas, TX 75080 USA (e-mail: aldhahir@utdallas.edu)}
\thanks{S. Zhang is with the Shenzhen Institute of Radio Testing and Tech, Shenzhen 518000 (e-mail: zhangsha@srtc.org.cn)}
\thanks{S. Li is with the School of Information
and Control Engineering, China University of Mining and Technology, Xuzhou
221116, China (e-mail: lishiyin@cumt.edu.cn).}

}

\maketitle
\begin{abstract}

 In this paper, we consider
a common unicast beamforming network where  Alice  utilizes the communication to Carol as a cover     and covertly transmits
  a message to Bob    without
being recognized by  Willie.
We investigate the beamformer design of Alice to maximize the covert rate to Bob when Alice has either perfect  or imperfect knowledge about Willie's channel state information (WCSI). For the perfect WCSI case, the problem is formulated under the perfect covert constraint, and
we develop a  covert beamformer by applying semidefinite relaxation  and the bisection method. Then, to reduce the computational complexity, we further
propose a  zero-forcing  beamformer design with a single  iteration processing.
For the case of  the imperfect WCSI,
the robust beamformer is developed based on a relaxation and restriction  approach by utilizing   the property of  Kullback-Leibler divergence.
Furthermore,  we derive the optimal   decision    threshold of  Willie, and analyze the false alarm and the missed detection probabilities in this case.
Finally, the performance of the proposed beamformer designs is evaluated through numerical experiments.

\end{abstract}

\IEEEpeerreviewmaketitle

\section{Introduction}


Due to its    broadcasting nature,
  wireless communication   is vulnerable to malicious attackers.
By exploiting encryption and key exchange
techniques,  conventional security methods
  mainly focus on preventing the transmitted wireless signals form
being decoded by unintended users \cite{Barros11Physical,Letafati20Lightweight}, but not concealing them. For many wireless scenarios, such as law enforcement and military communications, the transmitted signals should not be detected in order to perform the undercover missions.
Therefore,  the paradigm of covert communications, also known as   low probability
of detection (LPD) communications, aims to hide the transmissions status, and protect the users' privacy.

In a typical  covert communication scenario,  the sender (Alice) wants to send the information to the covert receiver (Bob) without being detected by the eavesdropper (Willie). Here, Willie may or may not be a legitimate receiver, while its purpose is to detect whether the transmission from Alice to Bob happens based on its observations.
 Mathematically,  the ultimate goal for Willie is to distinguish the two hypothesis ${{\cal H} _0}$ or  ${{\cal H} _1}$  by applying a specific decision rule, where    ${{\cal H} _0}$ denotes the null hypothesis  that
 Alice does not transmit a private  data stream to Bob, and 	${{\cal H} _1}$ denotes the alternate hypothesis that
Alice  transmits a private data stream to  Bob \cite{Lehmann_2005_Testing}.
In general,    the priori
probabilities of   hypotheses ${{{\cal H}_0}}$ and ${{{\cal H}_1}}$  are assumed to be equal, i.e., each equal to  $1/2$.
As such, the detection error probability
of  Willie  is defined as in \cite{Lehmann_2005_Testing,Yan2019Gaussian,Bash13}
\begin{align}\label{xi}
\xi  = \Pr \left( {{{\cal D}_1}\left| {{{\cal H}_0}} \right.} \right) + \Pr \left( {{{\cal D}_0}\left| {{{\cal H}_1}} \right.} \right) ,
\end{align}
where   ${{\cal D}_1}$ indicates  that Alice sends information to Bob, and  ${{\cal D}_0}$ indicates the other case.
Covert communication is achieved  for a given $\varepsilon \in \left[ {0,1} \right]$ if the detection error probability
$ \xi $ is no less than $1 - \varepsilon $, i.e., $ \xi  \ge 1 - \varepsilon $. Here, $\varepsilon $
is a predetermined value to specify the covert
communication constraint.

 Although practical covert communications
has been   studied  by investigating  spread-spectrum technology \cite{Simon94}
for several decades, the information-theoretic limits of covert communication
were only   recently derived \cite{Bash13,Bloch16,Wornell16}.
The achievability of the square root law (SRL) was established in \cite{Bash13}  where in order to achieve covert communication  over the  additive white Gaussian noise
(AWGN) channel, Alice can only transmit    no more than $\mathcal{O}\left(\sqrt n\right)$ bits  to Bob in
$n$ channel uses. Moreover, the  SRL  results  have been  verified     in  discrete memoryless
channels (DMCs) \cite{Wornell16,Bloch16}, two-hop systems\cite{Wu17},  multiple access channels \cite{Arumugam19} and broadcast channels \cite{Tan19}.
In short, these results imply that
the average number of covert bits per channel use asymptotically
approaches zero despite the noiseless transmission, i.e.,
 $\mathop {\lim }\limits_{n \to \infty } {{{\cal O}\left( {\sqrt n } \right)} \mathord{\left/
 {\vphantom {{{\mathcal O}\left( {\sqrt n } \right)} n}} \right.
 \kern-\nulldelimiterspace} n} = 0$.

Fortunately, some works have revealed that Alice can  achieve a positive
covert rate under some given conditions, i.e.,  imperfect knowledge of  noise  \cite{Lee15ach,Goeckel16cov,Yan17on};
   imperfect  channel statistics \cite{Che14rel,Shahzad2017Covert};  unknown transmission time
 \cite{BABash14LPD,Arumugam16Keyless,BABash16Covert}; the presence of random jamming signals \cite{Sobers2015Covert,Sobers2017Covert};  finite blocklengths of transmissions \cite{Yan19Delay,Huang2020LPD};
  the existence of relay \cite{Hu2018Covert,Forouzesh20Covert,Wang_WCL_2019};   and appearance of injecting artificial noise (AN) \cite{Soltani2018Covert,Shahzad2018Achieving}.
To be more specific,
   in  \cite{Forouzesh20Covert}, the authors proposed a power allocation strategy to maximize the secrecy rate under the covert
requirements  with  multiple untrusted     relays.
Based on the proposed rate-control and power-control strategies,  the authors in \cite{Hu2018Covert}   verified the feasibility of covert transmission   in amplify and forward one-way relay networks. With a finite number of channel uses,  delay-intolerant covert
communications was investigated in \cite{Yan19Delay}, which  demonstrated that
random transmit power  can enhance
covert communications.
In addition, the effect
of the  finite
blocklength (i.e., finite $n$) on covert communication was
investigated in \cite{Shahzad2018Achieving}. By exploiting a   full-duplex (FD) receiver,
covert communications was examined  in \cite{Shahzad2018Achieving} under fading channels, where the FD receiver
generates artificial noise   to confuse Willie.
In \cite{Yan2019Gaussian}, the optimality of Gaussian
signalling was investigated  by employing   Kullback-Leibler (KL) divergence as a covert  metric.
By formulating   LPD  communications   as
a quickest
detection problem,  the authors in \cite{Huang2020LPD} investigated  covert throughput maximization problems with three different  detection methods, i.e., the Shewhart, the cumulative sum (CUSUM),
and the Shiryaev-Roberts (SR) tests.
 With the help of a friendly uninformed jammer,    Alice
 can also
communicate  $\mathcal{O}\left( n\right)$ covert bits to Bob in $n$ channel uses \cite{Sobers2015Covert,Sobers2017Covert}.
 By producing artificial noise to inhibit Willie's detection, Alice can reliably and covertly transmit information to  Bob \cite{Soltani2018Covert}.

 Most existing works \cite{Bash13,Bloch16,Wornell16,Wu17,Arumugam19,Tan19,Sobers2015Covert,Sobers2017Covert,Hu2018Covert,Soltani2018Covert,Yan19Delay,Shahzad2018Achieving,Yan2019Gaussian,Huang2020LPD,Tao20Covert,Jiang20Covert} investigate  covert transmission with perfect channel state information (CSI) of all users.
However,
covert communication for multiple antenna beamforming  \cite{Forouzesh20Joint} has rarely been
studied to the best of our knowledge, and in practical
scenarios the perfect CSI of warden  is usually not available.
In \cite{Forouzesh20Joint}, the authors investigated  power allocation
 to maximize the secrecy rate while satisfying   the covert
communication requirements.
 In \cite{Forouzesh20Communication},      the  three-dimensional (3D) beamformer and jamming interference  beamformer were iteratively optimized for maximizing the covert rate.
In this work, we show that using multiple antennas allows us to relax the perfect CSI assumption while still guarantees   convert transmission.

 In this paper, we consider
a practical scenario where Alice uses the communication link with Carol as a cover, and  aims to achieve covert communication with Bob against Willie.
The most relevant work to this paper is \cite{Shahzad2017Covert}. In \cite{Shahzad2017Covert}, a single-input-single-output (SISO) covert communication scenario was considered, and an exact expression for the optimal threshold of Warden's detector was derived. The authors then analyzed the achievable rates with outage constraints under imperfect CSI.
However, in our work, we focus on a multiple-input-single-output (MISO) covert network for both the perfect CSI and   imperfect CSI cases.

 Our main contributions
are summarized as follows:

\begin{itemize}

\item When
Willie's  CSI (WCSI) is perfect at Alice, we study   the joint
  beamformer  design problem with the objective of maximizing the achievable rate of Bob, subject to the
perfect covert transmission constraint, quality of service (QoS) of Carol, and the total transmit power constraints
of Alice.
The covert rate maximization problem is shown to
be non-convex. Then, by applying the semidefinite
relaxation (SDR) technique and the bisection method, we find the solution by solving a series of convex  subproblems.

 \item Furthermore, to reduce the computational complexity, we propose a low-complexity zero-forcing (ZF) beamformer
design with a single iteration processing, which provides a promising tradeoff between complexity
and performance. Such design problem is transformed  into two decoupled subproblems. To be specific,
the design of  Willie's ZF beamformer can be relaxed to  a convex  problem by SDR, and Bob's ZF beamformer design can be reformulated as a convex second-order cone program (SOCP) problem.

\item When  WCSI is imperfect at Alice, we consider the robust covert rate maximization
problem   under the QoS constraint of Carol, the covertness constraint, and the total power constraint. To handle this non-convex problem, a restriction
and relaxation method is introduced, and a convex convex semidefinite program (SDP)
is obtained by using the S-lemma and SDR. Given that the covert constraint is not perfect, we derive the   optimal detection threshold   of Willie, and the corresponding detection error probability based on the robust beamformer vector. Such result can be used as the theoretical benchmark to evaluate the covert performance of   beamformers
deign.  Our simulation results further reveal the tradeoff between Willie's detection performance and Bob's covert rate.

 \end{itemize}

 The rest of this paper is organized as follows. In  Section II, we introduce the
system model, assumptions and the notations used throughout
the paper.
 In Section III, we discuss the covert beamformer and the ZF beamformer designs with perfect WCSI.
     In Section IV, we consider a robust beamforming design with imperfect WCSI.
        In Section V, we
  present numerical results to evaluate the proposed beamformers, and finally the paper
is concluded in Section   VI.

\emph{Notations}: Boldfaced lowercase and uppercase letters represent vectors and matrices, respectively.
 ${\mathop{\rm Re}\nolimits} \left(  \cdot  \right)$ and  ${\mathop{\rm Im}\nolimits} \left(  \cdot  \right)$
  denote the real part and imaginary part of its argument, respectively.
A complex-valued circularly symmetric Gaussian distribution with   mean   $\mu$ and  variance   ${\sigma ^2}$
is denoted by $\mathcal{CN}\left( {\mu ,{\sigma ^2}} \right)$.

\section{System Model}
\begin{figure}[h]
      \centering
	\includegraphics[width=8.5cm]{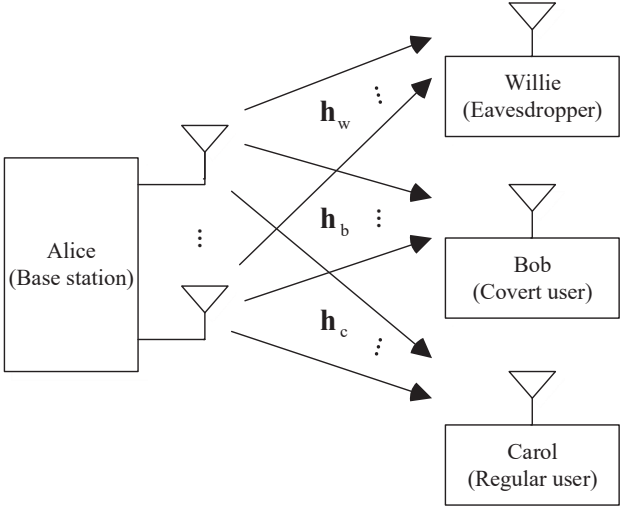}
 \caption{ Illustration of  the covert communication scenario}
  \label{VLC_RF} 
\end{figure}

We consider the scenario illustrated in Fig. 1, in which
Alice (base station) transmits  data stream ${{x_{{\rm{c}}}}}$  to    Carol (regular user)  all the time, and
 transmits private data stream ${x_{{\rm{b}}}}$  to    Bob (covert user) occasionally. For simplicity, let ${{\mathbb E}}\left\{ {{{\left| {{x_{\rm{c}}}} \right|}^2}} \right\} = 1$, ${{\mathbb E}}\left\{ {{{\left| {{x_{\rm{b}}}} \right|}^2}} \right\} = 1$.
  At the same time, Willie (eavesdropper) silently (passively) observes the communication environment and tries to identify whether Alice is transmitting to Bob or not.
 As we mentioned in the previous section, Alice may achieve covert communication by using the transmission to Carol as a cover.
 Suppose that  Alice is equipped with $N$ antennas, while
    Carol, Bob and Willie each has a single antenna\footnote{
    Under this setup, Willie only needs to perform energy detection and does not have to know the beamforming vectors.}.
 Let ${{\bf{h}}_{\rm{b}}} \in {\mathbb{C}^ N}$, ${{\bf{h}}_{\rm{c}}} \in {\mathbb{C}^N}$ and ${{\bf{h}}_{\rm{w}}} \in {\mathbb{C}^N}$
 denote the channel vectors from Alice to Bob, Carol and Willie, respectively.
We assume that all channels are modeled as Rayleigh flat fading, i.e.,
${{\bf{h}}_{\rm{b}}} \sim {\cal CN}\left( {{\bf{0}},\sigma_1^2{\bf{I}}} \right)$, ${{\bf{h}}_{\rm{w}}} \sim {\cal CN}\left( {{\bf{0}},\sigma_2^2{\bf{I}}} \right)$, and ${{\bf{h}}_{\rm{c}}} \sim {\cal CN}\left( {{\bf{0}},\sigma_3^2{\bf{I}}} \right)$ \cite{Shahzad2017Covert}, where $\sigma_1^2$, $\sigma_2^2$ and $\sigma_3^2$ denote the variances of   channels ${{\bf{h}}_{\rm{b}}}$, ${{\bf{h}}_{\rm{w}}}$ and ${{\bf{h}}_{\rm{c}}}$, respectively.

Recall that
the goal for Willie is to determine
which hypothesis  (${{\cal H} _0}$ or  ${{\cal H} _1}$) is true by applying a specific decision rule.
 For convenience,
we use ${{\cal D}_1}$  (${{\cal D}_0}$)  to indicate the event that  Alice does (does not) send information to Bob.

\subsection{Signal Model and Covert Constraints}
From Willie's perspective, Alice's transmitted signal   is given by
\begin{align}{\bf{x}} = \left\{ \begin{array}{l}
 {{\bf{w}}_{{\rm{c,0}}}}{x_{\rm{c}}},~~~~~~~~~~~{{\cal H} _0}, \\
 {{\bf{w}}_{{\rm{c,1}}}}{x_{\rm{c}}} + {{\bf{w}}_{\rm{b}}}{x_{\rm{b}}},~{{\cal H} _1}, \\
 \end{array} \right.
 \end{align}
where     ${{\bf{w}}_{\rm{c,0}}}$ and ${{\bf{w}}_{\rm{c,1}}}$ denote the  transmit beamformer vectors  for ${{x_{{\rm{c}}}}}$ in hypothesis  ${{\cal H} _0}$ and hypothesis  ${{\cal H} _1}$, respectively; ${{\bf{w}}_{\rm{b}}}$ denotes the  transmit beamformer vector for  ${{x_{{\rm{b}}}}}$. Let ${P_{{\rm{total}}}}$ denote the    maximum transmit power   of Alice. Therefore, the beamformer vectors satisfy: ${\left\| {{{\bf{w}}_{{\rm{c,0}}}}} \right\|^2} \le {P_{{\rm{total}}}}$ under ${{\cal H} _0}$ and ${\left\| {{{\bf{w}}_{{\rm{c,1}}}}} \right\|^2} + {\left\| {{{\bf{w}}_{\rm{b}}}} \right\|^2} \le {P_{{\rm{total}}}}$ under ${{\cal H} _1}$.

 For Carol, the received signal is given by
\begin{align}\label{yc}
{y_{{\rm{c}}}} =\left\{ {\begin{array}{*{20}{c}}
 {\bf{h}}_{\rm{c}}^H{{\bf{w}}_{{\rm{c}},0}}{x_{\rm{c}}} + {z_{\rm{c}}},~~~~~~~~~~~~~~{{\cal H} _0},\\
 {\bf{h}}_{\rm{c}}^H\left( {{{\bf{w}}_{{\rm{c}},{\rm{1}}}}{x_{\rm{c}}} + {{\bf{w}}_{\rm{b}}}{x_{\rm{b}}}} \right) + {z_{\rm{c}}},~{{\cal H} _1},
\end{array}} \right.
\end{align}
where   ${z_{\rm{c}}}\sim {\cal CN}\left( {{\rm{0}},\sigma _{\rm{c}}^2} \right)$ is  the received noise   at Carol\footnote{ Here, the inter cell interference  is  modeled as white Gaussian noise.}.

For Bob, the received signal is given by
\begin{align}\label{yb}
{y_{{\rm{b}}}}=\left\{ {\begin{array}{*{20}{c}}
{\bf{h}}_{\rm{b}}^H{{\bf{w}}_{{\rm{c}},0}}{x_{{\rm{c}}}} + {z_{\rm{b}}},~~~~~~~~~~~~~~{{\cal H} _0},\\
{\bf{h}}_{\rm{b}}^H\left( {{{\bf{w}}_{{\rm{c}},{\rm{1}}}}{x_{\rm{c}}} + {{\bf{w}}_{\rm{b}}}{x_{\rm{b}}}} \right) + {z_{\rm{b}}},~{{\cal H} _1},
\end{array}} \right.
\end{align}
where   ${z_{\rm{b}}}\sim {\cal CN}\left( {{\rm{0}},\sigma _{\rm{b}}^2} \right)$  is the received noise   at Bob.

In addition, the signals received by  Willie can be written as
\begin{align}\label{yw}
{y_{{\rm{w}}}}=\left\{ {\begin{array}{*{20}{c}}
{\bf{h}}_{\rm{w}}^H{{\bf{w}}_{{\rm{c}},0}}{x_{{\rm{c}}}} + {z_{\rm{w}}},~~~~~~~~~~~~~~{{\cal H} _0},\\
{\bf{h}}_{\rm{w}}^H\left( {{{\bf{w}}_{{\rm{c}},{\rm{1}}}}{x_{\rm{c}}} + {{\bf{w}}_{\rm{b}}}{x_{\rm{b}}}} \right) + {z_{\rm{w}}},~{{\cal H} _1},
\end{array}} \right.
\end{align}
where  ${z_{\rm{w}}}\sim {\cal CN}\left( {{\rm{0}},\sigma _{\rm{w}}^2} \right)$    is the received noise   at Willie.

According to \eqref{yc}, we assume that the instantaneous   rates at  Carol are expressed as  ${R_{{\rm{c}},0}}\left( {{{\bf{w}}_{{\rm{c}},0}}} \right) $ and ${R_{{\rm{c}},1}}\left( {{{\bf{w}}_{{\rm{c}},1}}},{{{\bf{w}}_{{\rm{b}}}}} \right)$ under ${{\cal H} _0}$ and ${{\cal H} _1}$,  respectively, and can be written as
\begin{subequations}\label{ratec}
\begin{align}
&{R_{{\rm{c}},0}}\left( {{{\bf{w}}_{{\rm{c}},0}}} \right) = \log_2 \left( {1 + \frac{{{{\left| {{\bf{h}}_{\rm{c}}^H{{\bf{w}}_{{\rm{c}},0}}} \right|}^2}}}{{\sigma _{\rm{c}}^2}}} \right),\\
 &{R_{{\rm{c}},1}}\left( {{{\bf{w}}_{{\rm{c}},1}}},{{{\bf{w}}_{{\rm{b}}}}} \right) = \log_2 \left( {1 + \frac{{{{\left| {{\bf{h}}_{\rm{c}}^H{{\bf{w}}_{{\rm{c}},1}}} \right|}^2}}}{{{{\left| {{\bf{h}}_{\rm{c}}^H{{\bf{w}}_{\rm{b}}}} \right|}^2} + \sigma _{\rm{c}}^2}}} \right).
 \end{align}
  \end{subequations}

Similarly, based on \eqref{yb}, we assume that ${R_{\rm{b}}}\left( {{{\bf{w}}_{{\rm{c}},1}}},{{{\bf{w}}_{{\rm{b}}}}} \right)$ is the instantaneous rate at  Bob  under hypothesis ${{\cal H} _1}$, which is  given by
\begin{align}\label{rateb}
{R_{\rm{b}}}\left( {{{\bf{w}}_{{\rm{c}},1}},{{\bf{w}}_{\rm{b}}}} \right) = {\log _2}\left( {1 + \frac{{{{\left| {{\bf{h}}_{\rm{b}}^H{{\bf{w}}_{\rm{b}}}} \right|}^2}}}{{{{\left| {{\bf{h}}_{\rm{b}}^H{{\bf{w}}_{{\rm{c}},1}}} \right|}^2} + \sigma _{\rm{b}}^2}}} \right).
\end{align}

Since Willie needs to distinguish between  the two hypotheses from its received signal ${y_{{\rm{w}}}}$, we further characterize the probability of ${y_{{\rm{w}}}}$.
Let  ${p_0}\left( {{y_{\rm{w}}}} \right)$
 and ${p_1}\left( {{y_{\rm{w}}}} \right)$
 denote the likelihood functions of  the received signals of Willie under ${{{\cal H}_{\rm{0}}}}$ and ${{{\cal H}_{\rm{1}}}}$, respectively.
Based on \eqref{yw}, ${p_0}\left( {{y_{\rm{w}}}} \right)$
 and ${p_1}\left( {{y_{\rm{w}}}} \right)$ are  given as
\begin{subequations}\label{C_8}
\begin{align}
& {p_0}\left( {{y_{\rm{w}}}} \right) = \frac{1}{{\pi {\lambda _0}}}\exp \left( { - \frac{{{{\left| {{y_{\rm{w}}}} \right|}^2}}}{{{\lambda _0}}}} \right), \\
 &{p_1}\left( {{y_{\rm{w}}}} \right) = \frac{1}{{\pi {\lambda _1}}}\exp \left( { - \frac{{{{\left| {{y_{\rm{w}}}} \right|}^2}}}{{{\lambda _1}}}} \right),
\end{align}
  \end{subequations}
where ${\lambda _0} \buildrel \Delta \over = {\left| {{\bf{h}}_{\rm{w}}^H{{\bf{w}}_{{\rm{c}},0}}} \right|^2} + \sigma _{\rm{w}}^2$ and $
{\lambda _1} \buildrel \Delta \over = {\left| {{\bf{h}}_{\rm{w}}^H{{\bf{w}}_{{\rm{c}},1}}} \right|^2} + {\left| {{\bf{h}}_{\rm{w}}^H{{\bf{w}}_{\rm{b}}}} \right|^2} + \sigma _{\rm{w}}^2$.

Recall from the previous section that Willie
 wants to   minimize the  detection error probability $ {\xi } $ \eqref{xi} by applying an optimal detector.
To take $\xi$ into our problem formulation, we next specify  conditions on the likelihood functions such that covert communication can be achieved with the given $\varepsilon$. First, we let
\begin{align}\label{C_7}
\xi
 = 1 - { V_T}\left( {{p_0},{p_1}} \right),
\end{align}
where ${V_T}\left( {{p_0},{p_1}} \right)$ is the total variation between ${p_0}\left( {{y_{\rm{w}}}} \right)$ and
${p_1}\left( {{y_{\rm{w}}}} \right)$.
In general, computing ${V_T}\left( {{p_0},{p_1}} \right)$
analytically is intractable. Thus, we adopt  Pinsker's inequality \cite{Cover_2003_Elements}, and can obtain
\begin{subequations}
\begin{align}
& { V_T}\left( {{p_0},{p_1}} \right) \le \sqrt {\frac{1}{2}D\left( {{p_0}\left\| {{p_1}} \right.} \right)},  \label{C_3}\\
 &{ V_T}\left( {{p_0},{p_1}} \right) \le \sqrt {\frac{1}{2}D\left( {{p_1}\left\| {{p_0}} \right.} \right)},\label{C_4}
\end{align}
  \end{subequations}
where $D\left( {{p_0}\left\| {{p_1}} \right.} \right)$ denotes the KL divergence from $p_0(y_{\rm{w}})$ to $p_1(y_{\rm{w}})$, and
$D\left( {{p_1}\left\| {{p_0}} \right.} \right)$ is the KL divergence from $p_1(y_{\rm{w}})$ to $p_0(y_{\rm{w}})$.
Furthermore,
$D\left( {{p_0}\left\| {{p_1}} \right.} \right)$   and
$D\left( {{p_1}\left\| {{p_0}} \right.} \right)$ are respectively given as
 \begin{subequations}
\begin{align}
  D\left( {{p_0}\left\| {{p_1}} \right.} \right) &= \int_{ - \infty }^{ + \infty } {{p_0}\left( {{y_{\text{w}}}} \right)\ln \frac{{{p_0}\left( {{y_{\text{w}}}} \right)}}{{{p_1}\left( {{y_{\text{w}}}} \right)}}} dy  = \ln \frac{{{\lambda _1}}}{{{\lambda _0}}} + \frac{{{\lambda _0}}}{{{\lambda _1}}} - 1, \label{C_5}\\
  D\left( {{p_1}\left\| {{p_0}} \right.} \right) &= \int_{ - \infty }^{ + \infty } {{p_1}\left( {{y_{\text{w}}}} \right)\ln \frac{{{p_1}\left( {{y_{\text{w}}}} \right)}}{{{p_0}\left( {{y_{\text{w}}}} \right)}}} dy = \ln \frac{{{\lambda _0}}}{{{\lambda _1}}} + \frac{{{\lambda _1}}}{{{\lambda _0}}} - 1. \label{C_6}
  \end{align}
\end{subequations}

Therefore,   to achieve  covert communication with the given $\varepsilon$, i.e., $ \xi    \ge 1 - \varepsilon $, the KL divergences  of the likelihood functions should satisfy  one of the following constraints:
\begin{subequations}\label{Dp0p1}
\begin{align}
&D\left( {{p_0}\left\| {{p_1}} \right.} \right) \le 2{\varepsilon ^2},\\
&D\left( {{p_1}\left\| {{p_0}} \right.} \right) \le 2{\varepsilon ^2}.
\end{align}
\end{subequations}

\subsection{CSI Availability}
In this subsection, we assume that Alice can accurately estimate the CSI of  Bob and Carol.
In most cases,
such CSI
can be learned at both the receiver side
and the transmitter side by training and feedback. However,
the WCSI    may not be always accessible to Alice because of the potential limited
cooperation between Alice and Willie. As a result,
  we consider
the following two  scenarios\footnote{
When Willie is totally passive, the covert communications scheme design may turn to exploit the channel distribution information  of Willie \cite{Zheng20Covert,Forouzesh20Joint,Forouzesh20Communication,Forouzesh19arXiv}
   }:

\textbf{1) Scenario 1. Perfect WCSI:} We first
consider a scenario that often arises in practice, where Willie
is a legitimate user and is only  hostile to Bob. In this case, Alice knows the full CSI  of the channel ${{\bf{h}}_{\rm{w}}}$, and uses it to help Bob to hide from Willie \cite{Bash13,Yan19Delay,Shahzad2018Achieving}.

\textbf{2) Scenario 2. Imperfect WCSI: }We consider a more practical scenario
where Willie is a regular user with only limited cooperation to Alice. In this case, Alice has imperfect CSI knowledge due to the passive warden and channel
estimation errors \cite{Shahzad2017Covert,Vakili06Effect}.
Here, the imperfect
   WCSI is modeled as
 \begin{align}
 {{\bf{h}}_{\rm{w}}} = {{{\bf{\hat h}}}_{\rm{w}}} + \Delta {{\bf{h}}_{\rm{w}}},
 \end{align}
 where ${{{\bf{\hat h}}}_{\rm{w}}}$  denotes the   estimated CSI vector between Alice and Willie, and $\Delta {{\bf{h}}_{\rm{w}}}$  denotes corresponding CSI error
vector. Moreover, the  CSI error vector $\Delta {{\bf{h}}_{\rm{w}}}$  is
characterized by an ellipsoidal region, i.e.,
\begin{align}
{{\cal E}_{\rm{w}}} \buildrel \Delta \over =\left\{ {\Delta {{\bf{h}}_{\rm{w}}}\left| {\Delta {\bf{h}}_{\rm{w}}^H{{\bf{C}}_{\rm{w}}}\Delta {{\bf{h}}_{\rm{w}}} \le {v_{\rm{w}}}} \right.} \right\}\label{C_25},
\end{align}
where ${{\bf{C}}_{\rm{w}}} = {{\bf{C}}_{\rm{w}}^H}\underline  \succ  {\bf{0}}$  controls the axes of the ellipsoid, and  ${v_{\rm{w}}}  >  0$  determines the volume of the ellipsoid \cite{BHe13Secure,Forouzesh20Communication}.

\section{Proposed  Covert Transmission for Perfect WCSI}

In this section, we consider the perfect WCSI scenario (scenario 1) and  maximize   the covert  rate to Bob by optimizing beamformers  at Alice. Specifically, we study a joint  beamforming  design problem with the objective of maximizing
  the achievable rate of Bob  ${R_{\rm{b}}}$, subject to the   perfect covert transmission constraint,  QoS of Carol, and the total transmit power constraints of Alice, which can be mathematically formulated as
\begin{subequations}\label{C_13}
\begin{align}
\mathop {\max }\limits_{{{\mathbf{w}}_{\rm{b}}},{{\mathbf{w}}_{{\rm{c}},{1}}}} {\rm{ }}&{R_{\rm{b}}}\left( {{{\bf{w}}_{{\rm{c}},1}}},{{{\bf{w}}_{{\rm{b}}}}} \right) \hfill \\
  {\text{s}}{\text{.t}}{\text{.}}~&{R_{{\rm{c}},1}}\left( {{{\bf{w}}_{{\rm{c}},1}}},{{{\bf{w}}_{{\rm{b}}}}} \right) = {R_{{\rm{c}},0}}\left( {{{\bf{w}}_{{\rm{c}},0}}} \right) , \hfill \\
  &D\left( {{p_0}\left\| {{p_1}} \right.} \right)= 0, \hfill \label{C_10}\\
 & {\left\| {{{\mathbf{w}}_{\rm{b}}}} \right\|^2}+{\left\| {{{\mathbf{w}}_{{\rm{c}},{1}}}} \right\|^2} \le  {P_{{\rm{total}}}}. \hfill
  \end{align}
\end{subequations}
Note that,   ${p_0}$ is a function of ${{\bf{w}}_{{\rm{c}},0}}$, and  ${p_1}$ is a function of both ${{\bf{w}}_{{\rm{c}},1}}$ and ${{\bf{w}}_{{\rm{b}}}}$ for \eqref{C_10}. Thus, ${p_0}$ and ${p_1}$ can be expressed as ${p_0}\left( {{{\bf{w}}_{{\rm{c}},0}}} \right)$ and ${p_1}\left( {{{\bf{w}}_{{\rm{c}},1}},{{\bf{w}}_{\rm{b}}}} \right)$, respectively.

  Notice that problem \eqref{C_13} is   non-convex and  difficult to be optimally solved.
Moreover,  constraints   $D\left( {{p_0}\left\| {{p_1}} \right.} \right)=0$ and $D\left( {{p_1}\left\| {{p_0}} \right.} \right)=0$ are equivalent for the  perfect covert transmission case.

 To address the non-convex   problem \eqref{C_13}, we  propose two beamformers design approaches, namely, the proposed covert beamformer design and proposed ZF beamformers design.

\subsection{Proposed Covert Beamformer Design}

 To simplify the derivation,  define
 $\tau _1  \buildrel \Delta \over ={\left| {{\mathbf{h}}_{{\rm{c}}}^H{{\mathbf{w}}_{{\rm{c}},0}}} \right|^2}$ and
  $\tau _2 \buildrel \Delta \over = {\left| {{\mathbf{h}}_{{\rm{w}}}^H{{\mathbf{w}}_{{\rm{c}},0}}} \right|^2}$,
   and introduce an auxiliary variable ${r_b}$.
   Then,
   problem \eqref{C_13} can be reformulated as the following equivalent form:
   \begin{subequations}\label{C_14}
\begin{align}
 \mathop {\max }\limits_{{{\bf{w}}_{\rm{b}}},{{\bf{w}}_{{\rm{c}},1}},{r_{\rm{b}}}} ~&{r_{\rm{b}}}\label{C_14a} \\
 {\rm{s.t.}}~&\frac{{{{\left| {{\bf{h}}_{\rm{b}}^H{{\bf{w}}_{\rm{b}}}} \right|}^2}}}{{{{\left| {{\bf{h}}_{\rm{b}}^H{{\bf{w}}_{{\rm{c}},1}}} \right|}^2} + \sigma _{\rm{b}}^2}} \ge {r_{\rm{b}}}, \\
 &\frac{{{{\left| {{\bf{h}}_{\rm{c}}^H{{\bf{w}}_{{\rm{c}},1}}} \right|}^2}}}{{{{\left| {{\bf{h}}_{\rm{c}}^H{{\bf{w}}_{\rm{b}}}} \right|}^2} + \sigma _{\rm{c}}^2}} = \frac{{{\tau _1}}}{{\sigma _{\rm{c}}^2}}, \\
 &{\left| {{\bf{h}}_{\rm{w}}^H{{\bf{w}}_{{\rm{c}},1}}} \right|^2} + {\left| {{\bf{h}}_{\rm{w}}^H{{\bf{w}}_{\rm{b}}}} \right|^2} = {\tau _2}, \\
 &{\left\| {{{\bf{w}}_{\rm{b}}}} \right\|^2} + {\left\| {{{\bf{w}}_{{\rm{c}},1}}} \right\|^2} \le {P_{{\rm{total}}}}.
 \end{align}
\end{subequations}

Next,  we apply the SDR technique \cite{ZLuo10Semidefinite} to relax  problem \eqref{C_14}. Towards this end, by using the following conditions
\begin{subequations}\label{SDR1}
\begin{align}
 &{{\bf{W}}_{\rm{b}}}{\rm{ = }}{{\bf{w}}_{\rm{b}}}{\bf{w}}_{\rm{b}}^H \Leftrightarrow {{\bf{W}}_{\rm{b}}} \underline  \succ  {\bf{0}},{\rm{rank}}\left( {{{\bf{W}}_{\rm{b}}}} \right) = 1, \\
 &{{\bf{W}}_{{\rm{c}},1}}{\rm{ = }}{{\bf{w}}_{{\rm{c}},1}}{\bf{w}}_{{\rm{c}},1}^H \Leftrightarrow {{\bf{W}}_{{\rm{c}},1}}\underline  \succ  {\bf{0}},{\rm{rank}}\left( {{{\bf{W}}_{{\rm{c}},1}}} \right) = 1,
\end{align}
\end{subequations}
   and ignoring the rank-one constraints,  we can obtain a relaxed version of problem \eqref{C_14}   as
\begin{subequations}\label{C_15}
\begin{align}
\mathop {\max }\limits_{{{\bf{W}}_{\rm{b}}},{{\bf{W}}_{{\rm{c}},1}},{r_{\rm{b}}}} &{r_{\rm{b}}}\\
{\rm{s}}.{\rm{t}}.~&{\rm{Tr}}\left( {{\bf{h}}_{\rm{b}}^H{{\bf{W}}_{\rm{b}}}{{\bf{h}}_{\rm{b}}}} \right) \ge {r_{\rm{b}}}\left( {{\rm{Tr}}\left( {{\bf{h}}_{\rm{b}}^H{{\bf{W}}_{{\rm{c}},1}}{{\bf{h}}_{\rm{b}}}} \right) + \sigma _{\rm{b}}^2} \right),\label{C_15a}\\
&\sigma _{\rm{c}}^2{\rm{Tr}}\left( {{\bf{h}}_{\rm{c}}^H{{\bf{W}}_{{\rm{c}},1}}{{\bf{h}}_{\rm{c}}}} \right) = {\tau _1}{\rm{Tr}}\left( {{\bf{h}}_{\rm{c}}^H{{\bf{W}}_{\rm{b}}}{{\bf{h}}_{\rm{c}}}} \right) + {\tau _1}\sigma _{\rm{c}}^2,\label{C_15b}\\
&{\rm{Tr}}\left( {{\bf{h}}_{\rm{w}}^H{{\bf{W}}_{{\rm{c}},1}}{{\bf{h}}_{\rm{w}}}} \right) + {\rm{Tr}}\left( {{\bf{h}}_{\rm{w}}^H{{\bf{W}}_{\rm{b}}}{{\bf{h}}_{\rm{w}}}} \right) = {\tau _2},\label{C_15c}\\
&{\rm{Tr}}\left( {{{\bf{W}}_{{\rm{c}},1}}} \right) + {\rm{Tr}}\left( {{{\bf{W}}_{\rm{b}}}} \right) \le {P_{{\rm{total}}}},\label{C_15d}\\
&{{{\bf{W}}_{{\rm{c}},1}}} \underline  \succ  {\bf{0}},~{{{\bf{W}}_{\rm{b}}}} \underline  \succ  {\bf{0}}.\label{C_15e}
\end{align}
\end{subequations}

Note that for any fixed ${r_{\rm{b}}} \ge 0$,
  problem \eqref{C_15} is a SDP. Therefore, problem \eqref{C_15} is quasi-concave, and
its optimal solution can be found by checking its feasibility   under any given ${r_{\rm{b}}}$.

 After that, it can be checked that the problem of maximizing \eqref{C_15a} is concave with respect to $r_{\rm{b}}$. To be more specific, let $\mathcal{W}=\{{{\bf{W}}_{\rm{b}}},{{\bf{W}}_{{\rm{c}},1}}|\eqref{C_15b}-\eqref{C_15e}\}$, $\phi({\bf{W}}_{\rm{b}}):={\rm{Tr}}\left( {{\bf{h}}_{\rm{b}}^H{{\bf{W}}_{\rm{b}}}{{\bf{h}}_{\rm{b}}}} \right)$, and $\theta({\bf{W}}_{{\rm{c}},1}):={{\rm{Tr}}\left( {{\bf{h}}_{\rm{b}}^H{{\bf{W}}_{{\rm{c}},1}}{{\bf{h}}_{\rm{b}}}} \right) + \sigma _{\rm{b}}^2}$. Then, we have the following result.

\textbf{Lemma 1:} {\it Function
\begin{align}
g(r_b)=\max_{\bf{W}_{\rm{b}},\bf{W}_{{\rm{c}},1}\in\mathcal{W}}& r_b,
\\
\textrm{s.t.}\quad&\phi({\bf{W}}_{\rm{b}})\ge r_b\theta({\bf{W}}_{{\rm{c}},1}).
\end{align}
 is concave for $r_b\ge0$.
}

\begin{IEEEproof}
Please see Appendix A.
\end{IEEEproof}

 Thus, we first  transform   problem \eqref{C_15}  into a series of convex
 subproblems with a given  ${r_{\rm{b}}} \ge 0$,  which can be optimally solved by standard convex optimization solvers
such as CVX. Next,
we adopt a bisection
search method to find the proposed covert  beamformers
${{\bf{W}}_{\rm{b}}}$ and ${{\bf{W}}_{{\rm{c}},1}}$. The details of  the bisection
search method  are summarized as Algorithm 1 in Table I, which outputs the   optimal solutions ${\bf{W}}_{{\rm{c}},1}^*$ and ${\bf{W}}_{\rm{b}}^*$.
  The   computational complexity of     Algorithm 1   is ${\cal O}\left( {\max {{\left\{ {4,2N} \right\}}^4}\sqrt {2N} \log \left( {{1 \mathord{\left/
 {\vphantom {1 {{\xi _1}}}} \right.
 \kern-\nulldelimiterspace} {{\xi _1}}}} \right)\log \left( {{1 \mathord{\left/
 {\vphantom {1 \zeta_1  }} \right.
 \kern-\nulldelimiterspace} \zeta_1  }} \right)} \right)$, where    ${\xi _1} > 0$ is the pre-defined accuracy of problem \eqref{C_15} \cite{ZLuo10Semidefinite,Grant09cvx,Sturm06SeDuMi}.

\begin{algorithm}[htb]
  \caption{Proposed covert beamformers design    method for   problem \eqref{C_15}}
  \label{alg:Bisection Method}
  \begin{algorithmic}[1]
    \State choose $\zeta_1   > 0$ (termination parameter), ${{ r}_{{\rm{b,l}}}}$ and ${{ r}_{{\rm{b,u}}}}$ such that ${ r}_{\rm{b}}^{\rm{*}}$ lies in $\left[ {{{ r}_{{\rm{b,l}}}},{{ r}_{{\rm{b,u}}}}} \right]$;
    \State Initialize ${{ r}_{{\rm{b,l}}}}=0$, ${{ r}_{{\rm{b,u}}}}={{\hat r}_{\rm{b}}}$;

     \While {${{ r}_{{\rm{b,u}}}} - {{ r}_{{\rm{b,l}}}} \ge \zeta_1  $}
    \State  set ${{ r}_{{\rm{b}}}} = \left( {{{ r}_{{\rm{b,l}}}} + {{ r}_{{\rm{b,u}}}}} \right)/2$;
    \State {\bf{if}}    problem \eqref{C_15}  is feasible, we get  solution ${{\bf{W}}_{\rm{b}}}$ and ${{\bf{W}}_{{\rm{c}},1}}$,  and set ${{ r}_{{\rm{b,l}}}} = {{ r}_{{\rm{b}}}}$
        \State {\bf{else}}, set ${{ r}_{{\rm{b,u}}}} = {{ r}_{{\rm{b,mid}}}}$;
    \EndWhile
  \State  Output   ${\bf{W}}_{{\rm{c}},1}^*$,${\bf{W}}_{\rm{b}}^*$;
  \end{algorithmic}
\end{algorithm}

Finally, we can  reconstruct the beamformers ${{\bf{w}}_{{\rm{c}},1}}$ and ${{\bf{w}}_{\rm{b}}}$ based on the solutions given by Algorithm 1.
Note that due to relaxation of SDR, the ranks of the optimal solutions ${\bf{W}}_{{\rm{c}},1}^*$,${\bf{W}}_{\rm{b}}^*$ may not be the optimal solutions of problem \eqref{C_13} or,  equivalently, \eqref{C_14}.
In particular, if ${\rm{rank}}\left( {{\bf{W}}_{{\rm{c}},1}^*} \right) = 1$ and ${\rm{rank}}\left( {{\bf{W}}_{\rm{b}}^*} \right) = 1$, then ${\bf{W}}_{{\rm{c}},1}^*$,${\bf{W}}_{\rm{b}}^*$ are also the optimal solutions of problem \eqref{C_13}, and the optimal beamformers ${{\bf{w}}_{{\rm{c}},1}}$ and ${{\bf{w}}_{\rm{b}}}$ can be obtained using the singular value decomposition (SVD), i.e.,${\bf{W}}_{{\rm{c}},1}^* = {{\bf{w}}_{{\rm{c}},1}}{\bf{w}}_{{\rm{c}},1}^H$ and ${\bf{W}}_{\rm{b}}^* = {{\bf{w}}_{\rm{b}}}{\bf{w}}_{\rm{b}}^H$.
  However, if ${\rm{rank}}\left( {{\bf{W}}_{{\rm{c}},1}^*} \right) > 1$ or ${\rm{rank}}\left( {{\bf{W}}_{\rm{b}}^*} \right) > 1$,
  we can adopt  the   Gaussian randomization procedure \cite{ZLuo10Semidefinite}  to produce a high-quality rank-one solution to problem \eqref{C_13}.

It is worth mentioning that  the above SDR based  beamformers design   approach requires   solving  a series of feasibility subproblems. Therefore,
this approach   leads to   high computational complexity, which motivates us
to further develop an alternative approach with   less intensive computational
complexity.

\subsection{Proposed Zero-Forcing Beamformers Design}

In this subsection, we propose  a ZF beamformers design  with iterative processing,  which is able to achieve a desirable tradeoff between complexity and
performance.
  In particular,
the interference signals ${\bf{h}}_{\rm{w}}^H{{\bf{w}}_{\rm{b}}}{s_{\rm{b}}}$ and  ${\bf{h}}_{\rm{c}}^H{{\bf{w}}_{\rm{b}}}{s_{\rm{b}}}$
 are eliminated by designing ${{\bf{w}}_{\rm{b}}}$
 such that  ${\bf{h}}_{\rm{w}}^H{{\bf{w}}_{\rm{b}}} = 0$ and  ${\bf{h}}_{\rm{c}}^H{{\bf{w}}_{\rm{b}}} = 0$.
 Meanwhile, the interference  signal  ${\bf{h}}_{\rm{b}}^H{{\bf{w}}_{{\rm{c}},1}}{s_{{\rm{c}},1}}$ can be removed
by designing ${{\bf{w}}_{{\rm{c}},1}}$
 such that  ${\bf{h}}_{\rm{b}}^H{{\bf{w}}_{{\rm{c}},1}}= 0$.
Note that ${{\bf{w}}_{\rm{b}}}$ has to be orthogonal to ${{\bf{h}}_{\rm{w}}} $ and ${{\bf{h}}_{\rm{c}}} $;
 and ${{\bf{w}}_{\rm{c,1}}}$ has to have non-zero projections
on ${{\bf{h}}_{\rm{w}}} $ and ${{\bf{h}}_{\rm{c}}} $, which means it has to be orthogonal to ${{\bf{h}}_{\rm{b}}} $. Since the
probability that ${{\bf{h}}_{\rm{b}}} $ falls in the space spanned by ${{\bf{h}}_{\rm{w}}} $ and ${{\bf{h}}_{\rm{c}}} $ is zero,
 the number of antennas at Alice is no less than three \cite{Jagannathan06Efficient,Kampeas18Ergodic}.

Mathematically, applying the ZF beamformers design   principle, problem \eqref{C_14} can be reformulated as
 \begin{subequations}\label{C_21}
\begin{align}
\mathop {\max }\limits_{{{\bf{w}}_{\rm{b}}},{{\bf{w}}_{\rm{c},1}}} {\rm{ }}{\rm{ }}{\rm{ }}{\rm{ }}& {\left| {{\bf{h}}_{\rm{b}}^H{{\bf{w}}_{\rm{b}}}} \right|^2}\label{C_21a}\\
{\rm{s.t.}}~&{\bf{h}}_{\rm{w}}^H{{\bf{w}}_{\rm{b}}} = 0,\label{C_23}\\
&{\bf{h}}_{\rm{c}}^H{{\bf{w}}_{\rm{b}}} = 0,\label{C_24}\\
&{\bf{h}}_{\rm{b}}^H{{\bf{w}}_{{\rm{c}},1}}= 0,\label{C_21c}\\
&{\left| {{\bf{h}}_{\rm{c}}^H{{\bf{w}}_{{\rm{c}},1}}} \right|^2} = {\tau _1},\label{C_21d}\\
&{\left| {{\bf{h}}_{\rm{w}}^H{{\bf{w}}_{{\rm{c}},1}}} \right|^2} = {\tau _2},\label{C_21e}\\
&{\left\| {{{\bf{w}}_{\rm{b}}}} \right\|^2} + {\left\| {{{\bf{w}}_{{\rm{c}},1}}} \right\|^2} \le {P_{\rm{{total}}}}\label{C_21g}.
 \end{align}
\end{subequations}

To address the joint ZF beamformers design   problem \eqref{C_21}, we first optimize the beamformer ${\bf{w}}_{\rm{c},1}$ by  minimizing the transmission power ${\left\| {{{\bf{w}}_{{\rm{c}},{\rm{1}}}}} \right\|^2}$
 under constraints \eqref{C_21c}, \eqref{C_21d} and \eqref{C_21e}.   This is because the objective function \eqref{C_21a}
 does not depend on ${{\bf{w}}_{\rm{c},1}}$, but it  increases with the power of beamformer  ${{\bf{w}}_{\rm{b}}}$.
 The total transmission power constraint \eqref{C_21g} includes  both ${{\bf{w}}_{\rm{b}}}$ and ${{\bf{w}}_{\rm{c},1}}$.
 Therefore, in order to maximize the objective function \eqref{C_21a}, we need to design  the beamformer ${\bf{w}}_{\rm{c},1}$ with the minimum  transmission power.
    Therefore, the ZF beamformer   ${\bf{w}}_{\rm{c},1}$ design   problem    can be formulated as
\begin{align}\label{C_21A}
\mathop {\min }\limits_{{{\bf{w}}_{{\rm{c}},1}}} {\rm{ }}&{\rm{ }}{\left\| {{{\bf{w}}_{{\rm{c}},1}}} \right\|^2}\\
{\rm{s.t.}}~&\eqref{C_21c}, \eqref{C_21d}, \eqref{C_21e},\nonumber
 \end{align}
which is also non-convex.

To handle the non-convexity issue, we  relax problem  \eqref{C_21A} to a convex form by applying SDR as well, which is similar to the approach we followed in the previous section.
Specifically, by relaxing ${{\bf{W}}_{{\rm{c}},1}} = {{\bf{w}}_{{\rm{c}},1}}{\bf{w}}_{{\rm{c}},1}^H$
to ${{\bf{W}}_{{\rm{c}},1}} \succeq {\bf{0}}$, problem  \eqref{C_21A} can be reformulated as
 \begin{subequations}\label{C_21A2}
\begin{align}
\mathop {\min }\limits_{{{\bf{W}}_{{\rm{c}},1}}}{\rm{ }} &{\rm{ }}{\rm{Tr}}\left( {{{\bf{W}}_{{\rm{c}},1}}} \right)\\
 {\rm{s.t.}}~&{\rm{Tr}}\left( {{{\bf{W}}_{{\rm{c}},1}}{{\bf{h}}_{\rm{b}}}{\bf{h}}_{\rm{b}}^H} \right) = 0, \\
 &{\rm{Tr}}\left( {{{\bf{W}}_{{\rm{c}},1}}{{\bf{h}}_{\rm{c}}}{\bf{h}}_{\rm{c}}^H} \right) = {\tau _1}, \\
 &{\rm{Tr}}\left( {{{\bf{W}}_{{\rm{c}},1}}{{\bf{h}}_{\rm{w}}}{\bf{h}}_{\rm{w}}^H} \right) = {\tau _2},\\
 &{{\bf{W}}_{{\rm{c}},1}} \succeq {\bf{0}},
 \end{align}
\end{subequations}
which is a convex SDP.

Let ${\bf{W}}_{{\rm{c}},1}^{{\rm{opt}}}$  denote the optimal solution of problem \eqref{C_21A2}.
Due to relaxation, the rank of  ${\bf{W}}_{{\rm{c}},1}^{{\rm{opt}}}$  may not equal to one.
Therefore if ${\rm{rank}}\left( {{\bf{W}}_{{\rm{c}},1}^{{\rm{opt}}}} \right) = 1$, then ${\bf{W}}_{{\rm{c}},1}^{{\rm{opt}}}$ is the optimal solution of  problem \eqref{C_21}, and the optimal beamformer  ${{\bf{w}}_{{\rm{c}},1}}$   can be obtained by SVD, i.e.,${\bf{W}}_{{\rm{c}},1}^{{\rm{opt}}} = {{\bf{w}}_{{\rm{c}},1}}{\bf{w}}_{{\rm{c}},1}^H$.
Otherwise, if ${\rm{rank}}\left( {{\bf{W}}_{{\rm{c}},1}^{{\rm{opt}}}} \right) > 1$,
  we can adopt  the   Gaussian randomization procedure \cite{ZLuo10Semidefinite}  to produce a high-quality rank-one solution to problem \eqref{C_21A}.

Next, we consider the design of ${{\bf{w}}_{{\rm{b}}}}$.  Let ${{\bf{w}}_{{\rm{c}},{\rm{1}}}^{{\rm{opt}}}}$ denote the   beamformer of problem \eqref{C_21A2}.
Then, let
${P_c}={\left\| {{\bf{w}}_{{\rm{c}},{\rm{1}}}^{{\rm{opt}}}} \right\|^2}$ denote the transmission power of ${{\bf{w}}_{{\rm{c}},{\rm{1}}}^{{\rm{opt}}}}$. With the notations just defined,   problem \eqref{C_21} can be formulated as
 \begin{subequations}\label{C_21B}
\begin{align}
\mathop {\max }\limits_{{{\bf{w}}_{\rm{b}}}}{\rm{ }}& {\rm{ }}{\left| {{\bf{h}}_{\rm{b}}^H{{\bf{w}}_{\rm{b}}}} \right|^2}\\
{\rm{s.t.}}~&{\left\| {{{\bf{w}}_{\rm{b}}}} \right\|^2} + {P_c} \le {P_{{\rm{total}}}},\label{C_21Ba}\\
&\eqref{C_23}, \eqref{C_24},\notag
 \end{align}
\end{subequations}
which is equivalent to
\begin{subequations}\label{C_22}
\begin{align}
\mathop {\max }\limits_{{{\bf{w}}_{\rm{b}}}} &~{\mathop{\rm Re}\nolimits} \left\{ {{\bf{h}}_{\rm{b}}^H{{\bf{w}}_{\rm{b}}}} \right\}\\
{\rm{s}}{\rm{.t}}{\rm{.}}~&{\rm{Im}}\left\{ {{\bf{h}}_{\rm{b}}^H{{\bf{w}}_{\rm{b}}}}  \right\} = 0,\\
& \eqref{C_23}, \eqref{C_24}, \eqref{C_21Ba}.\nonumber
 \end{align}
\end{subequations}

Problem \eqref{C_22} is   a  SOCP that can be optimally solved by standard convex optimization solvers
such as CVX \cite{Grant09cvx}. Therefore, the ZF   transmit beamformers   of problem
\eqref{C_21} are finally obtained.

Furthermore, we analyze       the multiplexing gains  of the covert communication system based on ZF beamformer design \cite{Kampeas18Ergodic}. Specifically, based on the   definitions    ${{\bf{H}}_{{\rm{w,c}}}} \buildrel \Delta \over = \left[ {{{\bf{h}}_{\rm{w}}},{{\bf{h}}_{\rm{c}}}} \right]$ and $\prod _{{\rm{w,c}}}^ \bot  \buildrel \Delta \over = {\bf{I}} - \frac{{{{\bf{H}}_{{\rm{w,c}}}}{\bf{H}}_{{\rm{w,c}}}^H}}{{{{\left\| {{{\bf{H}}_{{\rm{w,c}}}}} \right\|}^2}}}$, the ZF beamformer ${{\bf{w}}_{\rm{b}}}$  of   problem \eqref{C_24} is given as
    \begin{align}{{\bf{w}}_{\rm{b}}} = \alpha \frac{{\prod _{{\rm{w,c}}}^ \bot {{\bf{h}}_{\rm{b}}}}}{{\left\| {\prod _{{\rm{w,c}}}^ \bot {{\bf{h}}_{\rm{b}}}} \right\|}},
     \end{align}
     where  $\alpha  \ge 0$ is a non-negative real-valued scalar.
     Then, by
     substituting ${{\bf{w}}_{\rm{b}}}$ into problem \eqref{C_24},   the optimal ZF beamformer of Bob is given by
       \begin{align}{{\bf{w}}_{\rm{b}}} = \sqrt {{P_{{\rm{total}}}} - {P_c}} \frac{{\prod _{{\rm{w,c}}}^ \bot {{\bf{h}}_{\rm{b}}}}}{{\left\| {\prod _{{\rm{w,c}}}^ \bot {{\bf{h}}_{\rm{b}}}} \right\|}}.
         \end{align}

 Thus, the   multiplexing gain  covert communication system is given as
\begin{subequations}\label{multiplexing}
\begin{align}
&\mathop {\lim }\limits_{{P_{{\rm{total}}}} \to \infty } \frac{{{R_b}}}{{{\rm{lo}}{{\rm{g}}_2}{\rm{SNR}}}}\\
 &= \mathop {\lim }\limits_{\alpha  \to \infty } \frac{{{{\log }_2}\left( {\frac{{{P_{{\rm{total}}}} - {P_c}}}{{\sigma _{\rm{b}}^2}}} \right) + {{\log }_2}{{\left| {\frac{{{\bf{h}}_{\rm{b}}^H\prod _{{\rm{w,c}}}^ \bot {{\bf{h}}_{\rm{b}}}}}{{\left\| {\prod _{{\rm{w,c}}}^ \bot {{\bf{h}}_{\rm{b}}}} \right\|}}} \right|}^2}}}{{{\rm{lo}}{{\rm{g}}_2}\frac{{{P_{{\rm{total}}}} - {P_c}}}{{\sigma _{\rm{b}}^2}}}}\\
 &= 1,
  \end{align}
\end{subequations}
 where ${\rm{SNR}} \buildrel \Delta \over = \frac{{{{\left\| {{{\bf{w}}_{\rm{b}}}} \right\|}^2}}}{{\sigma _{\rm{b}}^2}}$.

 \section{Proposed Robust  Covert Transmission for Imperfect WCSI}

In the previous section, we considered the   case with perfect  WCSI. In practice, it is common that the obtained CSI is corrupted by certain estimation errors \cite{Wornell16,Bloch16}. Hence,  we further propose a robust beamforming
design for the optimization problem \eqref{C_13}  under the imperfect WCSI  scenario.
In this scenario, the perfect covert transmission, i.e., $D\left( {{p_0}\left\| {{p_1}} \right.} \right) = 0$, is difficult to achieve. Therefore,
 we adopt     $D\left( {{p_0}\left\| {{p_1}} \right.} \right) \le 2{\varepsilon ^2}$ and
$D\left( {{p_1}\left\| {{p_0}} \right.} \right) \le 2{\varepsilon ^2}$ as covertness constraints\cite{Bash13,Wornell16,Bloch16,Yan2019Gaussian},
according to \eqref{Dp0p1}. Moreover, based on the developed robust beamformer, we further study the best situation for Willie where the desired detection error probability of Willie can be achieved.

\subsection{ Case of $D\left( {{p_0}\left\| {{p_1}} \right.} \right) \le 2{\varepsilon ^2}$}

With imperfect WCSI, we aim to maximize ${R_b}$ via the joint design of the beamformers ${{\bf{w}}_{{\rm{c}},1}}$ and ${{\bf{w}}_{\rm{b}}}$
  under the  QoS  of Carol, the covertness constraint and the total power constraint.
 Mathematically,  the robust covert rate maximization problem  is formulated as
\begin{subequations}\label{problem3}
\begin{align}
\mathop {\max }\limits_{{{\mathbf{w}}_{\rm{b}}},{{\mathbf{w}}_{{\rm{c}},{1}}}} {\rm{ }}~&{R_{\rm{b}}}\left( {{{\bf{w}}_{{\rm{c}},1}}},{{{\bf{w}}_{{\rm{b}}}}} \right) \hfill \\
  {\text{s}}{\text{.t}}{\text{.}}~&{R_{{\rm{c}},1}}\left( {{{\bf{w}}_{{\rm{c}},1}}},{{{\bf{w}}_{{\rm{b}}}}} \right) = {R_{{\rm{c}},0}}\left( {{{\bf{w}}_{{\rm{c}},0}}} \right) , \hfill \label{problem3a}\\
  &D\left( {{p_0}\left\| {{p_1}} \right.} \right) \le 2{\varepsilon ^2}, \hfill \label{problem3b}\\
 & {\left\| {{{\mathbf{w}}_{\rm{b}}}} \right\|^2}+{\left\| {{{\mathbf{w}}_{{\rm{c}},{1}}}} \right\|^2} \le  {P_{{\rm{total}}}}, \label{problem3c}\\
 &{{\bf{h}}_{\rm{w}}} = {{{\bf{\hat h}}}_{\rm{w}}} + \Delta {{\bf{h}}_{\rm{w}}}, \Delta {{\bf{h}}_{\rm{w}}} \in {{\cal{E}}_{\rm{w}}}. \label{problem3d}
  \end{align}
\end{subequations}
Recall in Section II.B that the CSIs of Bob  and Carol, i.e., ${{\bf{h}}_{\rm{b}}}$ and ${{\bf{h}}_{\rm{c}}}$,  are perfectly known.

It is clear that  problem \eqref{problem3} is not convex, and thereby it is difficult
to obtain the optimal solution directly.
To deal with
this issue, we first reformulate the  covertness constraint \eqref{problem3c},
 by exploiting the property  of the function $f\left( x \right) = \ln x + \frac{1}{x} - 1$ for $x > 0$.
More specifically,    the covertness constraint $D\left( {{p_0}\left\| {{p_1}} \right.} \right) = \ln \frac{{{\lambda _1}}}{{{\lambda _0}}} + \frac{{{\lambda _0}}}{{{\lambda _1}}} - 1 \le 2{\varepsilon ^2}$ can be  equivalently transformed as
 \begin{align}
\bar a \le \frac{{{\lambda _1}}}{{{\lambda _0}}} \le \bar b\label{barab},
 \end{align}
where    $\bar a$ and $\bar b$ are the two roots of the equation $\ln \frac{{{\lambda _1}}}{{{\lambda _0}}} + \frac{{{\lambda _0}}}{{{\lambda _1}}} - 1 = 2{\varepsilon ^2}$.
Therefore, constraint \eqref{problem3b} can   be equivalently reformulated as
\begin{align}
&\bar a \le \frac{{{{\left| {{\bf{h}}_{\rm{w}}^H{{\bf{w}}_{{\rm{c}},1}}} \right|}^2} + {{\left| {{\bf{h}}_{\rm{w}}^H{{\bf{w}}_{\rm{b}}}} \right|}^2} + \sigma _{\rm{w}}^2}}{{{{\left| {{\bf{h}}_{\rm{w}}^H{{\bf{w}}_{\rm{c,0}}}} \right|}^2} + \sigma _{\rm{w}}^2}} \le \bar b.\label{C_39}
 \end{align}

Here, due to $\Delta {{\bf{h}}_{\rm{w}}} \in {{\cal{E}}_{\rm{w}}}$, there are infinite choices for
$\Delta {{\bf{h}}_{\rm{w}}}$, in constraint \eqref{problem3d},
 which makes problem \eqref{problem3} non-convex and intractable.
To overcome this challenge, we propose a   relaxation and restriction approach. Specifically,
in the relaxation step, the nonconvex
robust design problem is transformed into a convex
SDP; while in
the restriction step, infinite number of complicated
constraints are reformulated into a finite number of linear matrix
inequalities (LMIs).

For mathematical convenience, let us define ${{\bf{W}}_{\rm{b}}}  =  {{\bf{w}}_{\rm{b}}}{{\bf{w}}_{\rm{b}}}^H$,    ${{\bf{W}}_{{\rm{c}},1}}  =  {{\bf{w}}_{{\rm{c}},1}}{\bf{w}}_{{\rm{c}},1}^H$, $\widehat {\bf{W}}_1 \buildrel \Delta \over = {{\bf{W}}_b} + {{\bf{W}}_{c,1}} - \bar a{{\bf{w}}_{{\rm{c}},0}}{\bf{w}}_{{\rm{c}},0}^H$, and $\widetilde {\bf{W}}_1 \buildrel \Delta \over = {{\bf{W}}_{{\rm{c}},1}} + {{\bf{W}}_{\rm{b}}}- \bar b{{\bf{w}}_{{\rm{c}},0}}{\bf{w}}_{{\rm{c}},0}^H$.
Then,  constraint \eqref{C_39} can be  equivalently re-expressed as
\begin{subequations}
\begin{align}
&\Delta {\bf{h}}_{\rm{w}}^H\widehat {\bf{W}}_1\Delta {{\bf{h}}_{\rm{w}}} + 2\Delta {\bf{h}}_{\rm{w}}^H\widehat {\bf{W}}_1{{{\bf{\hat h}}}_{\rm{w}}} + {\bf{\hat h}}_{\rm{w}}^H\widehat {\bf{W}}_1{{{\bf{\hat h}}}_{\rm{w}}} \ge \sigma _{\rm{w}}^2\left( {\bar a - 1} \right)\label{W1a},\\
& \Delta {\bf{h}}_{\rm{w}}^H\widetilde {\bf{W}}_1\Delta {{\bf{h}}_{\rm{w}}} + 2\Delta {\bf{h}}_{\rm{w}}^H\widetilde {\bf{W}}_1{{{\bf{\hat h}}}_{\rm{w}}} + {\bf{\hat h}}_{\rm{w}}^H\widetilde {\bf{W}}_1{{{\bf{\hat h}}}_{\rm{w}}}\le \sigma _{\rm{w}}^2\left( {\bar b - 1} \right)\label{W1b},
 \end{align}
 \end{subequations}

By applying SDR, we  ignore  the rank-one constraints of  ${{\bf{W}}_{{\rm{c}},1}}$ and ${{\bf{W}}_{\rm{b}}}$,
   which is similar to the approach used in \eqref{SDR1} and  \eqref{C_15}. Then,   problem \eqref{problem3}  can be relaxed as follows
\begin{subequations}\label{C_41}
\begin{align}
\mathop {\max }\limits_{{{\bf{W}}_{\rm{b}}},{{\bf{W}}_{{\rm{c}},1}},{{\tilde r}_{\rm{b}}}} &{{\tilde r}_{\rm{b}}}\\
{\rm{s}}.{\rm{t}}.~&{\rm{Tr}}\left( {{\bf{h}}_{\rm{b}}^H{{\bf{W}}_{\rm{b}}}{{\bf{h}}_{\rm{b}}}} \right) \ge {{\tilde r}_{\rm{b}}}\left( {{\rm{Tr}}\left( {{\bf{h}}_{\rm{b}}^H{{\bf{W}}_{{\rm{c}},1}}{{\bf{h}}_{\rm{b}}}} \right) + \sigma _{\rm{b}}^2} \right)\label{C_41a},\\
&\sigma _{\rm{c}}^2{\rm{Tr}}\left( {{\bf{h}}_{\rm{c}}^H{{\bf{W}}_{{\rm{c}},1}}{{\bf{h}}_{\rm{c}}}} \right) = {\tau _1}{\rm{Tr}}\left( {{\bf{h}}_{\rm{c}}^H{{\bf{W}}_{\rm{b}}}{{\bf{h}}_{\rm{c}}}} \right) + {\tau _1}\sigma _{\rm{c}}^2,\label{C_41b}\\
&{\rm{Tr}}\left( {{{\bf{W}}_{{\rm{c}},1}}} \right) + {\rm{Tr}}\left( {{{\bf{W}}_{\rm{b}}}} \right) \le {P_{{\rm{total}}}},\label{C_41d}\\
&{{{\bf{W}}_{{\rm{c}},1}}} \underline  \succ  {\bf{0}},~{{{\bf{W}}_{\rm{b}}}} \underline  \succ  {\bf{0}},\label{C_41e}\\
&{\Delta {\bf{h}}_{\rm{w}}^H{{\bf{C}}_{\rm{w}}}\Delta {{\bf{h}}_{\rm{w}}} \le {v_w}},\\
&\eqref{W1a}, \eqref{W1b}\notag.
 \end{align}
\end{subequations}
where ${{\tilde r}_{\rm{b}}} \ge 0$ is a slack variable.

Note that the SDR problem \eqref{C_41} is quasi-concave, since the objective
function and constraints are linear in ${{\bf{W}}_{{\rm{c}},1}}$ and ${{\bf{W}}_{\rm{b}}}$. However, problem \eqref{C_41}
is still computationally intractable because it involves an infinite
number of constraints due to $\Delta {{\bf{h}}_{\rm{w}}} \in {{\cal E}_{\rm{w}}}$.

 Next, we employ the S-Procedure to recast the infinitely many constraints
  as a certain set of LMIs, which  is a tractable  approximation.

\textbf{Lemma 2 (S-Procedure\cite{DWKRobust14})}: Let a function ${f_m}\left( x \right),m \in \left\{ {1,2} \right\},x \in {{\mathbb{C}}^{N \times 1}}$, be defined as
\begin{align}
{f_m}\left( x \right) = {{\bf{x}}^H}{{\bf{A}}_m}{\bf{x}} + 2\text{Re}\left\{ {{\bf{b}}_m^{\rm{H}}{\bf{x}}} \right\} + {c_m},
\end{align}
where ${{\bf{A}}_m} \in {{\mathbb{C}}^N}$ is a complex
Hermitian matrix, ${{\bf{b}}_m} \in {{\mathbb{C}}^{N \times 1}}$ and ${c_m} \in {{\mathbb{R}}^{1 \times 1}}$. Then, the
implication relation ${f_1}\left( x \right) \le 0 \Rightarrow {f_2}\left( x \right) \le 0$ holds if and  only if there
exists a variable $\eta   \ge 0$  such that
\begin{align}
{\eta } \left[ {\begin{array}{*{20}{c}}
{{{\bf{A}}_1}}&{{{\bf{b}}_1}}\\
{{\bf{b}}_1^{{H}}}&{{c_1}}
\end{array}} \right] - \left[ {\begin{array}{*{20}{c}}
{{{\bf{A}}_2}}&{{{\bf{b}}_2}}\\
{{\bf{b}}_2^{{H}}}&{{c_2}}
\end{array}} \right]\underline  \succ  {\bf{0}}\label{spro}.
\end{align}

Consequently,
by using the S-Procedure of Lemma 2,  constraints \eqref{W1a} and \eqref{W1b} can be respectively recast  as a finite number of LMIs:
\begin{subequations}
\begin{align}
\left[ {\begin{array}{*{20}{c}}
{\widehat {\bf{W}}_1 + {\eta _1}{{\bf{C}}_{\rm{w}}}}&{\widehat {\bf{W}}_1{{{\bf{\hat h}}}_{\rm{w}}}}\\
{{\bf{\hat h}}_{\rm{w}}^H\widehat {\bf{W}}_1}&{{\bf{\hat h}}_{\rm{w}}^H\widehat {\bf{W}}_1{{{\bf{\hat h}}}_{\rm{w}}} - \sigma _{\rm{w}}^2\left( {\bar a - 1} \right) - {\eta _1}{v_{\rm{w}}}}
\end{array}} \right]\underline  \succ  {\bf{0}}\label{slemma1},\\
\left[ {\begin{array}{*{20}{c}}
{ - \widetilde {\bf{W}}_1 + {\eta _2}{{\bf{C}}_{\rm{w}}}}&{ - \widetilde {\bf{W}}_1{{{\bf{\hat h}}}_{\rm{w}}}}\\
{ - {\bf{\hat h}}_{\rm{w}}^H\widetilde {\bf{W}}_1}&{ - {\bf{\hat h}}_{\rm{w}}^H\widetilde {\bf{W}}_1{{{\bf{\hat h}}}_{\rm{w}}} + \sigma _{\rm{w}}^2\left( {\bar b - 1} \right)   - {\eta _2}{v_{\rm{w}}}}
\end{array}} \right]\underline  \succ  {\bf{0}}\label{slemma2}.
\end{align}
\end{subequations}

Thus, we obtain the following conservative approximation of problem \eqref{C_41}:
\begin{align}\label{C_42}
\mathop {\max }\limits_{{{\bf{W}}_{\rm{b}}},{{\bf{W}}_{{\rm{c}},1}},{{\tilde r}_{\rm{b}}}} {\rm{ }}&{{\tilde r}_{\rm{b}}}\\
{\rm{s}}.{\rm{t}}.~&\eqref{C_41a}, \eqref{C_41b}, \eqref{C_41d},\eqref{C_41e}, \eqref{slemma1}, \eqref{slemma2}\notag.
 \end{align}

When ${{\tilde r}_{\rm{b}}}$ is fixed, problem  \eqref{C_42} is a convex SDP which can be efficiently solved
by off-the-shelf convex solvers \cite{Grant09cvx}. Therefore,  problem  \eqref{C_42} can be efficiently solved by  the proposed bisection method, which is summarized in Algorithm  2.
  The   computational complexity of     Algorithm 2   is ${\cal O}\left( {\max {{\left\{ {5,2N - 1} \right\}}^4}\sqrt {2N - 1} \log \left( {{1 \mathord{\left/
 {\vphantom {1 {{\xi _2}}}} \right.
 \kern-\nulldelimiterspace} {{\xi _2}}}} \right)\log \left( {{1 \mathord{\left/
 {\vphantom {1 \zeta_2  }} \right.
 \kern-\nulldelimiterspace} {\zeta _2} }} \right)} \right)$, where    ${\xi _2} > 0$ is the pre-defined accuracy of problem \eqref{C_42}.

Similarly,   if ${\rm{rank}}\left( {{\bf{W}}_{{\rm{c}},1}^*} \right) = 1$ and ${\rm{rank}}\left( {{\bf{W}}_{\rm{b}}^*} \right) = 1$, then ${\bf{W}}_{{\rm{c}},1}^*$,${\bf{W}}_{\rm{b}}^*$ are also the optimal solutions of problem \eqref{problem3}, and the optimal beamformers ${{\bf{w}}_{{\rm{c}},1}}$ and ${{\bf{w}}_{\rm{b}}}$ can be obtained by SVD, i.e.,${\bf{W}}_{{\rm{c}},1}^* = {{\bf{w}}_{{\rm{c}},1}}{\bf{w}}_{{\rm{c}},1}^H$ and ${\bf{W}}_{\rm{b}}^* = {{\bf{w}}_{\rm{b}}}{\bf{w}}_{\rm{b}}^H$.
  However, if ${\rm{rank}}\left( {{\bf{W}}_{{\rm{c}},1}^*} \right) > 1$ or ${\rm{rank}}\left( {{\bf{W}}_{\rm{b}}^*} \right) > 1$,
  we can adopt  the   Gaussian randomization procedure \cite{ZLuo10Semidefinite}  to produce a high-quality rank-one solution to problem \eqref{problem3}.

\begin{algorithm}[htb]
  \caption{Proposed robust beamformers design method for   problem  \eqref{C_42}}
  \label{alg:2}
  \begin{algorithmic}[1]
    \State choose $\zeta_2  > 0$ (termination parameter), ${{\tilde r}_{{\rm{b,l}}}}$ and ${{\tilde r}_{{\rm{b,u}}}}$ such that ${\tilde r}_{\rm{b}}^{\rm{*}}$ lies in $\left[ {{{\tilde r}_{{\rm{b,l}}}},{{\tilde r}_{{\rm{b,u}}}}} \right]$;
    \State Initialize ${{\tilde r}_{{\rm{b,l}}}}=0$, ${{\tilde r}_{{\rm{b,u}}}}={{\hat r}_{\rm{b}}}$;
     \While {${{\tilde r}_{{\rm{b,u}}}} - {{\tilde r}_{{\rm{b,l}}}} \ge \zeta_2$}
    \State  Let ${{\tilde r}_{{\rm{b}}}} = \left( {{{\tilde r}_{{\rm{b,l}}}} + {{\tilde r}_{{\rm{b,u}}}}} \right)/2$;
    \State  {\bf{if}}    problem \eqref{C_42} is feasible, we obtain the   solution ${{\bf{W}}_{\rm{b}}}$ and ${{\bf{W}}_{{\rm{c}},1}}$, and set ${{\tilde r}_{{\rm{b,l}}}} = {{\tilde r}_{{\rm{b}}}}$;
         \State {\bf{else}}, let ${{\tilde r}_{{\rm{b,u}}}} = {{\tilde r}_{{\rm{b}}}}$;
    \EndWhile
   \State Output the optimal solutions ${\bf{W}}_{{\rm{c}},1}^*$,${\bf{W}}_{\rm{b}}^*$.
  \end{algorithmic}
\end{algorithm}

\subsection{Case of $D\left( {{p_1}\left\| {{p_0}} \right.} \right) \le 2{\varepsilon ^2}$}

In this subsection, we consider the constraint $D\left( {{p_1}\left\| {{p_0}} \right.} \right) \le 2{\varepsilon ^2}$, and the   corresponding robust covert rate maximization problem  can be   formulated as
\begin{subequations}\label{problem4}
\begin{align}
\mathop {\max }\limits_{{{\mathbf{w}}_{\rm{b}}},{{\mathbf{w}}_{{\rm{c}},{1}}}} {\rm{ }}&{R_{\rm{b}}}\left( {{{\bf{w}}_{{\rm{c}},1}}},{{{\bf{w}}_{{\rm{b}}}}} \right) \hfill \\
  {\text{s}}{\text{.t}}{\text{.}}~&{R_{{\rm{c}},1}}\left( {{{\bf{w}}_{{\rm{c}},1}}},{{{\bf{w}}_{{\rm{b}}}}} \right) = {R_{{\rm{c}},0}}\left( {{{\bf{w}}_{{\rm{c}},0}}} \right) , \hfill \\
  &D\left( {{p_1}\left\| {{p_0}} \right.} \right) \le 2{\varepsilon ^2}, \hfill \label{problem4b}\\
  & {\left\| {{{\mathbf{w}}_{\rm{b}}}} \right\|^2}+{\left\| {{{\mathbf{w}}_{{\rm{c}},{1}}}} \right\|^2} \le  {P_{{\rm{total}}}}, \hfill\\
 &{{\bf{h}}_{\rm{w}}} = {{{\bf{\hat h}}}_{\rm{w}}} + \Delta {{\bf{h}}_{\rm{w}}}, \Delta {{\bf{h}}_{\rm{w}}} \in {{\cal{E}}_{\rm{w}}} \label{problem4d},
  \end{align}
\end{subequations}
where $D\left( {{p_1}\left\| {{p_0}} \right.} \right) = \ln \frac{{{\lambda _0}}}{{{\lambda _1}}} + \frac{{{\lambda _1}}}{{{\lambda _0}}} - 1$.

Note that problem \eqref{problem4}
 is similar to problem \eqref{problem3} except for the covertness constraint.
The covertness constraint $D\left( {{p_1}\left\| {{p_0}} \right.} \right) = \ln \frac{{{\lambda _0}}}{{{\lambda _1}}} + \frac{{{\lambda _1}}}{{{\lambda _0}}} - 1 \le 2{\varepsilon ^2}$ can be  equivalently transformed as
\begin{align}
\bar c \le \frac{{{\lambda _0}}}{{{\lambda _1}}} \le \bar d,
 \end{align}
 where $\bar c= {\bar a}$ and $\bar d= {\bar b}$, are the two roots of the equation $\ln \frac{{{\lambda _0}}}{{{\lambda _1}}} + \frac{{{\lambda _1}}}{{{\lambda _0}}} - 1 = 2{\varepsilon ^2}$.

Similar to the previous subsection, we may apply the relaxation and restriction approach to solve problem \eqref{problem4}. We omit the detailed derivations for brevity. Note that although the methods are similar, the achievable covert rates  are quite different under the two covertness constraints. We will illustrate and discuss this issue in   the next section.

\subsection{Ideal Detection Performance  of Willie}

In order to evaluate the above robust beamformer deign, we further develop the optimal decision threshold of Willie, and the corresponding  false alarm
 and missed detection probabilities. We consider  the ideal case for Willie, i.e., the beamformers
${\mathbf{w}}_{\rm{b}}$, ${{\mathbf{w}}_{{\rm{c}},{0}}}$, and ${{\mathbf{w}}_{{\rm{c}},{1}}}$ are   known by Willie, which is the
    the worst case for Bob.

 According to the
Neyman-Pearson criterion \cite{Lehmann_2005_Testing}, the optimal  rule  for Willie to
minimize his  detection error is the   likelihood
ratio test\cite{Lehmann_2005_Testing}, i.e.,
 \begin{align}\label{criterion}
\frac{{{p_1}\left( {{y_{\rm{w}}}} \right)}}{{{p_0}\left( {{y_{\rm{w}}}} \right)}}\frac{{\mathop  > \limits^{{D_1}} }}{{\mathop  < \limits_{{D_0}} }}1,
\end{align}
where
${{{\cal D}_1}}$ and ${{{\cal D}_0}}$ are the  binary decisions that correspond to hypotheses ${{{\cal H}_0}}$
and  ${{{\cal H}_1}}$, respectively.
Furthermore, \eqref{criterion} can be equivalently reformulated as
\begin{align}
{\left| {{y_{\rm{w}}}} \right|^2} \frac{{\mathop  > \limits^{{{\cal D}_1}} }}{{\mathop  < \limits_{{{\cal D}_0}} }} {\phi ^*}.
\end{align}
where  ${\phi ^*} \buildrel \Delta \over = \frac{{{\lambda _0}{\lambda _1}}}{{{\lambda _1} - {\lambda _0}}}{\rm{ln}}\frac{{{\lambda _1}}}{{{\lambda _0}}}$ denotes the optimal detection threshold of Willie.
Here, please recall that $\lambda _0$ and $\lambda _1$ are given in \eqref{C_8}, which depend on the beamformer vectors ${\mathbf{w}}_{\rm{b}}$, ${{\mathbf{w}}_{{\rm{c}},{0}}}$, and ${{\mathbf{w}}_{{\rm{c}},{1}}}$.

According to \eqref{C_8},   the cumulative density functions (CDFs) of ${\left| {{y_{\rm{w}}}} \right|^2}$ under ${{{\cal H}_{\rm{0}}}}$ and ${{{\cal H}_{\rm{1}}}}$ are respectively given by
\begin{subequations}\label{C_9}
\begin{align}
 &\Pr \left( {{{\left| {{y_{\rm{w}}}} \right|}^2}|{\cal{H}_{\rm{0}}}} \right) = 1 - \exp \left( { - \frac{{{{\left| {{y_{\rm{w}}}} \right|}^2}}}{{{\lambda _0}}}} \right), \\
 &\Pr \left( {{{\left| {{y_{\rm{w}}}} \right|}^2}|{\cal{H}_{\rm{1}}}} \right) = 1 - \exp \left( { - \frac{{{{\left| {{y_{\rm{w}}}} \right|}^2}}}{{{\lambda _1}}}} \right).
 \end{align}
\end{subequations}
Therefore,  based on   the optimal detection threshold ${\phi ^*}$, the false alarm $ P \left( {{{\cal D}_1}\left| {{{\cal H}_0}} \right.} \right)$ and   missed detection probabilities   $P \left( {{{\cal D}_0}\left| {{{\cal H}_1}} \right.} \right)$  are given as
\begin{subequations}
\begin{align}
 &P \left( {{{\cal D}_1}\left| {{{\cal H}_0}} \right.} \right)= {\Pr}\left( {{{\left| {{y_{\rm{w}}}} \right|}^2} \ge {\phi ^*}|{{{\cal H}_{\rm{0}}}}} \right) = {\left( {\frac{{{\lambda _1}}}{{{\lambda _0}}}} \right)^{ - \frac{{{\lambda _1}}}{{{\lambda _1} - {\lambda _0}}}}}, \\
 &P \left( {{{\cal D}_0}\left| {{{\cal H}_1}} \right.} \right) = {\Pr}\left( {{{\left| {{y_{\rm{w}}}} \right|}^2} \le {\phi ^*}|{{{\cal H}_{\rm{1}}}}} \right)  = 1 - {\left( {\frac{{{\lambda _1}}}{{{\lambda _0}}}} \right)^{ - \frac{{{\lambda _0}}}{{{\lambda _1} - {\lambda _0}}}}}.
\end{align}
\end{subequations}

Therefore, the ideal detection performance of Willie can be characterized by  ${\phi ^*}$, $P \left( {{{\cal D}_1}\left| {{{\cal H}_0}} \right.} \right)$ and $P \left( {{{\cal D}_0}\left| {{{\cal H}_1}} \right.} \right)$  . Such results can be used as
the theoretical benchmark to evaluate the covert performance
of our robust beamformer designs. We will further discuss the detection performance of Willie in the next section.

\section{Numerical Results}

In this section, we present and discuss  numerical results  to assess the performance of the proposed
covert beamformers design, ZF beamformers design and  robust beamformers design methods for covert communications.
In our simulations, we set  the number of   antennas  at  Alice to $5$, i.e., $N=5$,   the noise
variance of the three  users is    normalized to $1$,
i.e., $\sigma _{\rm{c}}^2=\sigma _{\rm{b}}^2=\sigma _{\rm{w}}^2=0\rm{dBW}$, the total transmit power of Alice to $P_{\rm{total}}=10\rm{dBW}$, and ${{{{\left\| {{{\bf{w}}_{c,0}}} \right\|}^2}}}=1\rm{dBW}$.
 Moreover, we assume that all channels experience Rayleigh flat fading, and $\sigma _1=\sigma _2=\sigma _3=1$ \cite{Shahzad2017Covert}.

\subsection{Evaluation for Scenario 1}
 We first evaluate the proposed methods in scenario 1, i.e., Alice with perfect WCSI.
 \begin{figure}[h]
          \centering
      \includegraphics[width=7.5cm]{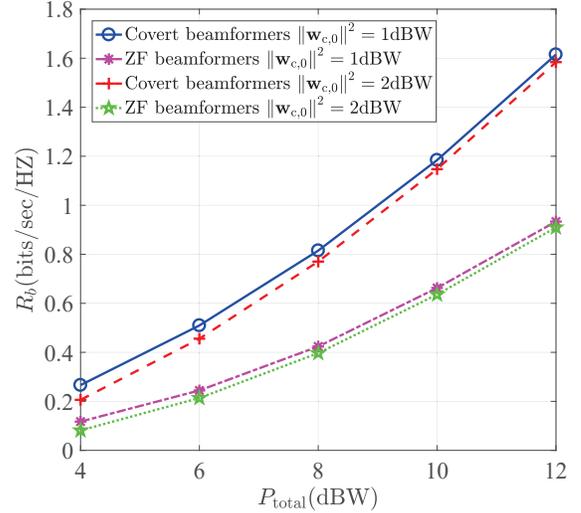}
 \caption{~$R_{\rm{b}}$ (bits/sec/Hz) of the proposed covert beamformer design and proposed ZF beamformer design design versus $P_{\rm{total}}$ (dBW).}
    \label{Ptotal_rb} 
\end{figure}

 Fig. \ref{Ptotal_rb} depicts the covert rate of Bob $R_{\rm{b}}$ with the proposed covert beamformer design and the proposed ZF beamformer design versus the total transmit power $P_{\rm {total}}$.
 It can be observed that the covert rate of Bob $R_{\rm{b}}$ increases as
the transmit power of Alice $P_{\rm {total}}$ increases, while $R_{\rm{b}}$ of the proposed covert beamforming design is higher than
that of the ZF beamformer design. In addition, by comparing the two different transmit powers of beamformer for  Carol ${{{{\left\| {{{\bf{w}}_{{\rm{c}},0}}} \right\|}^2}}}$ under ${\cal {H}}_0$, we observe that the lower  the transmit power of ${{{{\left\| {{{\bf{w}}_{{\rm{c}},0}}} \right\|}^2}}}$ is, the higher the covert rate of Bob $R_{\rm{b}}$ will be.
  This is because  when the transmit power of  ${{{{\left\| {{{\bf{w}}_{{\rm{c}},0}}} \right\|}^2}}}$ is lower,
  more power can be allocated to Bob.

\begin{figure}[h]
      \centering
	\includegraphics[width=7.5cm]{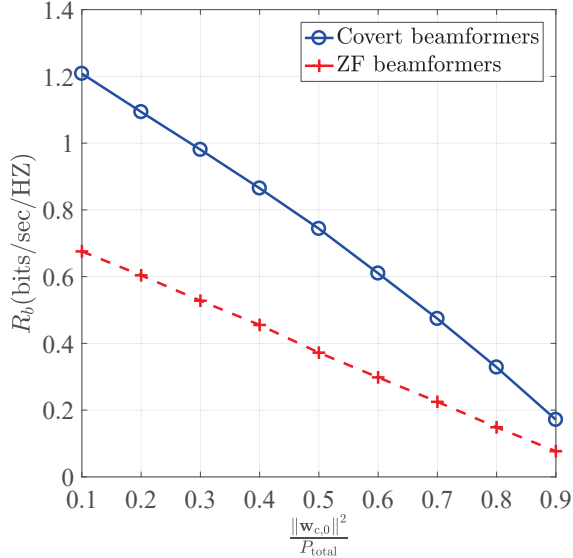}
 \caption{~$R_{\rm{b}}$ of the proposed covert beamformer design and proposed ZF beamformer design versus
  with different ratios $\frac{{{{\left\| {{{\bf{w}}_{{\rm{c}},0}}} \right\|}^2}}}{P_{{\rm{total}}}}$.}
  \label{percent} 
\end{figure}

 Fig. \ref{percent}  plots the covert rate     $R_{\rm{b}}$ of the proposed covert beamformer design and the proposed ZF beamformer design versus
    different ratios $\frac{{{{\left\| {{{\bf{w}}_{{\rm{c}},0}}} \right\|}^2}}}{P_{{\rm{total}}}}$ with ${P_{{\rm{total}}}}=10 \rm{W}$.
In this figure, we
observe that for a fixed value of the ratio $\frac{{{{\left\| {{{\bf{w}}_{{\rm{c}},0}}} \right\|}^2}}}{P_{{\rm{total}}}}$,   $R_{\rm{b}}$ of the ZF beamformer design  is lower than that  of the covert beamformer design, which   is consistent with  Fig. \ref{Ptotal_rb}.
 In addition,
as the ratio $\frac{{{{\left\| {{{\bf{w}}_{{\rm{c}},0}}} \right\|}^2}}}{P_{{\rm{total}}}}$ increases,  the covert rate of Bob $R_{\rm{b}}$   decreases, and the rate gap between the covert beamformer design and ZF beamformer design   also decreases.
This is because when the ratio $\frac{{{{\left\| {{{\bf{w}}_{{\rm{c}},0}}} \right\|}^2}}}{P_{{\rm{total}}}}$ is high, the allocated power  of   the beamformer $ {{{\bf{w}}_{{\rm{b}}}}}$ is  close to $0$, which leads to rate gap between the covert beamformer design  and   ZF design close to $0$; and when the ratio   is low, more power is allocated to the beamformer  $ {{{\bf{w}}_{{\rm{b}}}}}$, which results in a
 rate gap between   the covert beamformer design  and   ZF design (verified in Fig. \ref{Ptotal_rb}).

\begin{figure}[h]
      \centering
	\includegraphics[width=7.5cm]{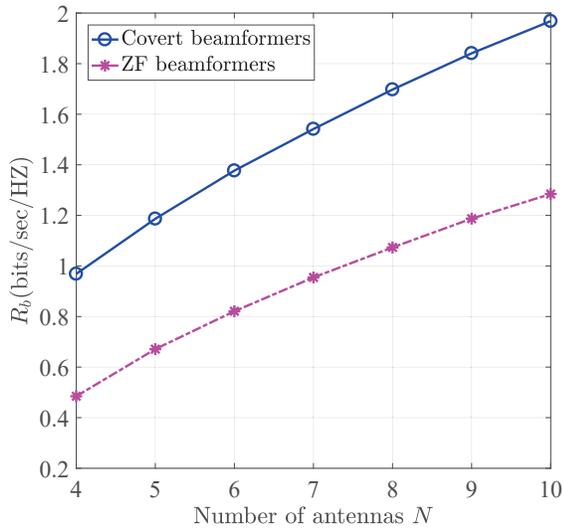}
 \caption{ $R_{\rm{b}}$ versus the number of  antennas $N$ for proposed covert beamformer design and proposed ZF beamformer design.}
  \label{N_rb} 
\end{figure}

In Fig.  \ref{N_rb}, we plot the covert rate of Bob $R_{\rm{b}}$ of the proposed covert beamformer design and the proposed ZF beamformer design versus
  the   number of antennas of Alice $N$    with ${P_{{\rm{total}}}}=10 \rm{dBW}$.
It is observed that  as the number of antennas $N$ increases, the covert rate of Bob $R_{\rm{b}}$ increases  and  the rate gap between the covert beamformer design and ZF beamformer design   also increases. This is because with more  antennas,
 more   spatial multiplexing gains can be exploited.

 Through   Figs. \ref{Ptotal_rb}, \ref{percent}  and \ref{N_rb}, we observe that the  covert rate of the proposed covert beamformer design is always higher than that of the proposed ZF beamformer design.
 However, the computational complexity of the ZF beamformer design is significantly  lower than  that of the covert beamformer design.
  Specifically, the comparison of the computational time between the covert beamformer design and ZF beamformer design is presented in Table I, and all simulations of the two methods are performed using MATLAB 2016b with 2.30GHz, 2.29GHz dual CPUs and a 128GB RAM.
  Table I shows that  the computational time  of the covert beamformer design and ZF beamformer design  increases as the number of antennas $N$ increases.
  More importantly,
  the computational time of the ZF beamformer design is less than ${1 \mathord{\left/
 {\vphantom {1 {10}}} \right.
 \kern-\nulldelimiterspace} {10}}$
 of that of the covert beamformer design.
\begin{table}[htbp]
  \centering
  \caption{Comparison of the computational time between the proposed covert beamformer design and proposed ZF beamformer design }
  \begin{tabular}{|c|c|c|c|c|}
   \hline
    \diagbox{Method}{Time/second}{$N$}&$N=4$  &$N=6$ &$N=8$ &$N=10$ \\
   \hline
Covert Design&$10.36$&$10.49$&$10.97$&$11.39$\\
\hline
ZF Design&$0.7571$&$0.7593$&$0.7615$&$0.7621$\\
\hline
  \end{tabular}
 \end{table}

\subsection{Evaluation for Scenario 2}
In this subsection, we evaluate the proposed robust beamformer design for scenario 2, namely, Alice with imperfect WCSI.

\begin{figure}
    \begin{minipage}[b]{0.45\textwidth}
      \centering
      \includegraphics[height=7.5cm,width=7.5cm]{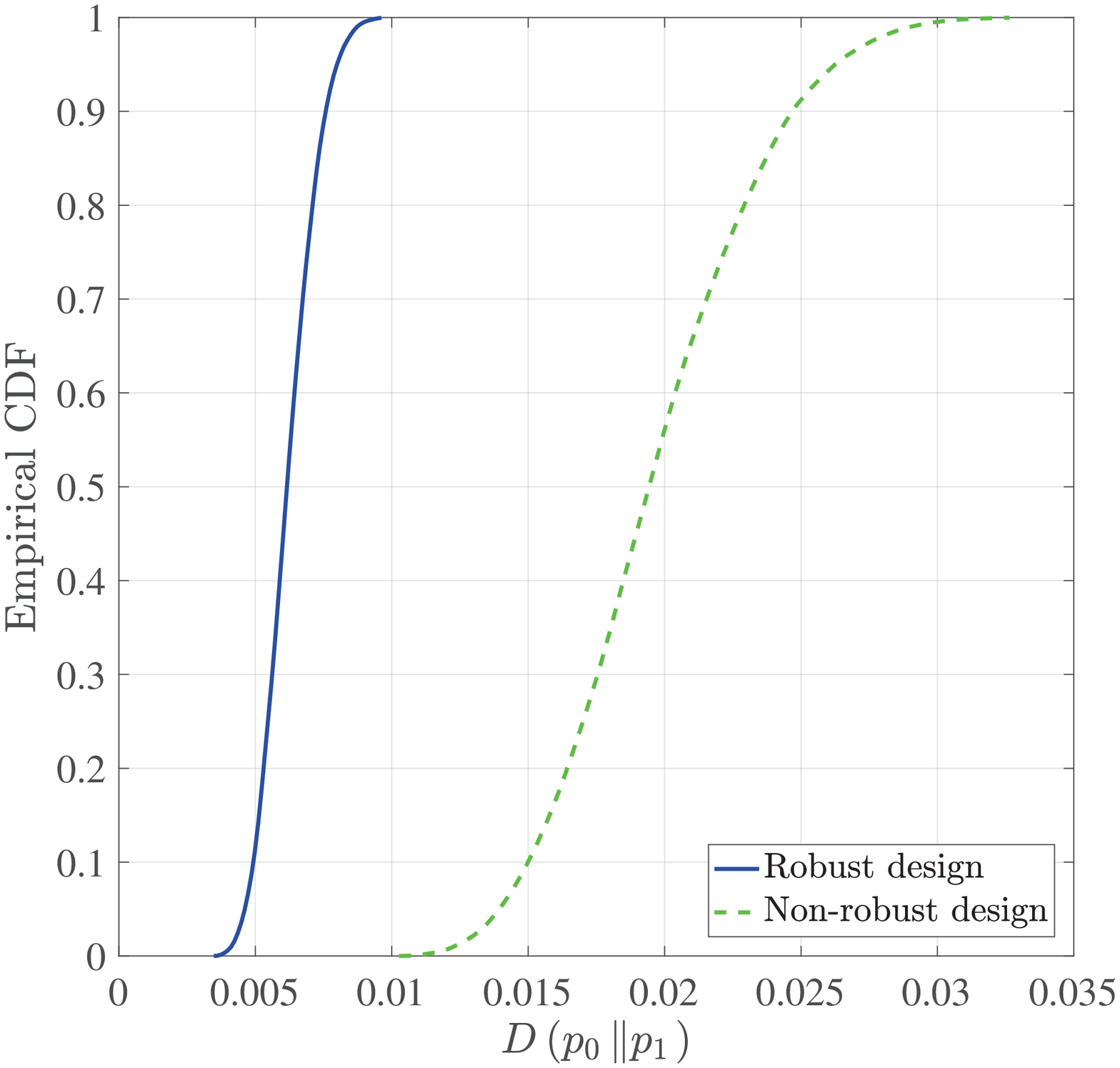}
      \vskip-0.2cm\centering {\footnotesize (a)}
    \end{minipage}
     \begin{minipage}[b]{0.45\textwidth}
      \centering
      \includegraphics[height=7.5cm,width=7.5cm]{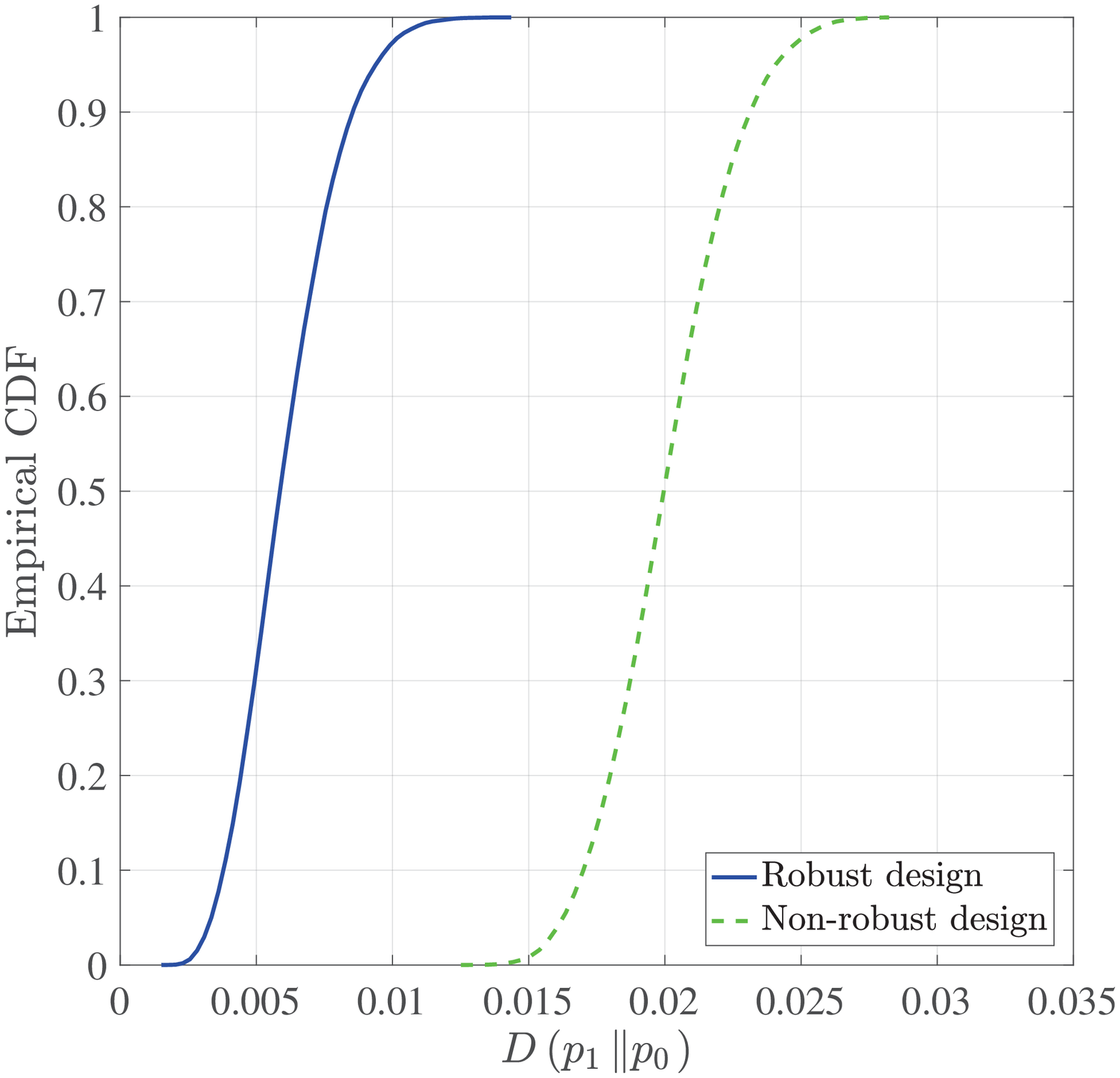}
      \vskip-0.2cm\centering {\footnotesize (b)}
    \end{minipage}\hfill
 \caption{The empirical CDF of (a) $D\left( {{p_0}\left\| {{p_1}} \right.} \right)$  and (b) $D\left( {{p_0}\left\| {{p_1}} \right.} \right)$, with the covertness threshold   $2{\varepsilon ^2} = 0.02$ and  CSI errors   $v_w=0.005$.}
 \label{cdf}  
\end{figure}

  Fig. \ref{cdf} shows the cumulative density function (CDF) of $D\left( {{p_0}\left\| {{p_1}} \right.} \right) $, where  the relative entropy requirement is $D\left( {{p_0}\left\| {{p_1}} \right.} \right) \le 0.02$,  ${{{{\left\| {{{\bf{w}}_{{\rm{c}},0}}} \right\|}^2}}}=8 {\rm{dBW}}$ and $v_w=0.005$.
From these results, we observe that the CDF in the KL divergence of the non-robust design cannot guarantee the requirement, while the robust beamforming design satisfies the KL divergence constraint, that is, it satisfies Willie's error detection probability requirement, in order to achieve our goal.

Fig. \ref{cdf} (a) and (b) show the empirical CDF  of the achieved $D\left( {{p_0}\left\| {{p_1}} \right.} \right)$ and $D\left( {{p_1}\left\| {{p_0}} \right.} \right)$, respectively, for both the robust  and non-robust designs, where the covertness threshold is $2{\varepsilon ^2} = 0.02$, i.e., $D\left( {{p_0}\left\| {{p_1}} \right.} \right) \le 0.02$ and $D\left( {{p_1}\left\| {{p_0}} \right.} \right) \le 0.02$, and the CSI errors parameter is $v_w=0.005$.
Here, the non-robust design refers to the proposed covert design with ${{{\bf{\hat h}}}_{\rm{w}}} $ under the same conditions.
    As can be
observed from  Fig. \ref{cdf} (a) and (b), the proposed robust design satisfies the covertness
constraint,  i.e., $D\left( {{p_0}\left\| {{p_1}} \right.} \right) \le 0.02$ and $D\left( {{p_1}\left\| {{p_0}} \right.} \right) \le 0.02$. On the other hand, the
non-robust design cannot satisfy the  covertness
constraints, where
about 45$\%$ of the resulting $D\left( {{p_0}\left\| {{p_1}} \right.} \right)$   exceed   the covertness
threshold $2{\varepsilon ^2} = 0.02$; and about 50$\%$ of the resulting $D\left( {{p_1}\left\| {{p_0}} \right.} \right)$   exceed   the covertness
threshold $2{\varepsilon ^2} = 0.02$. Fig. \ref{cdf} (a) and (b) verify     the necessity and effectiveness of the proposed robust design.

\begin{figure}
    \begin{minipage}[b]{0.45\textwidth}
      \centering
      \includegraphics[height=7.5cm,width=7.5cm]{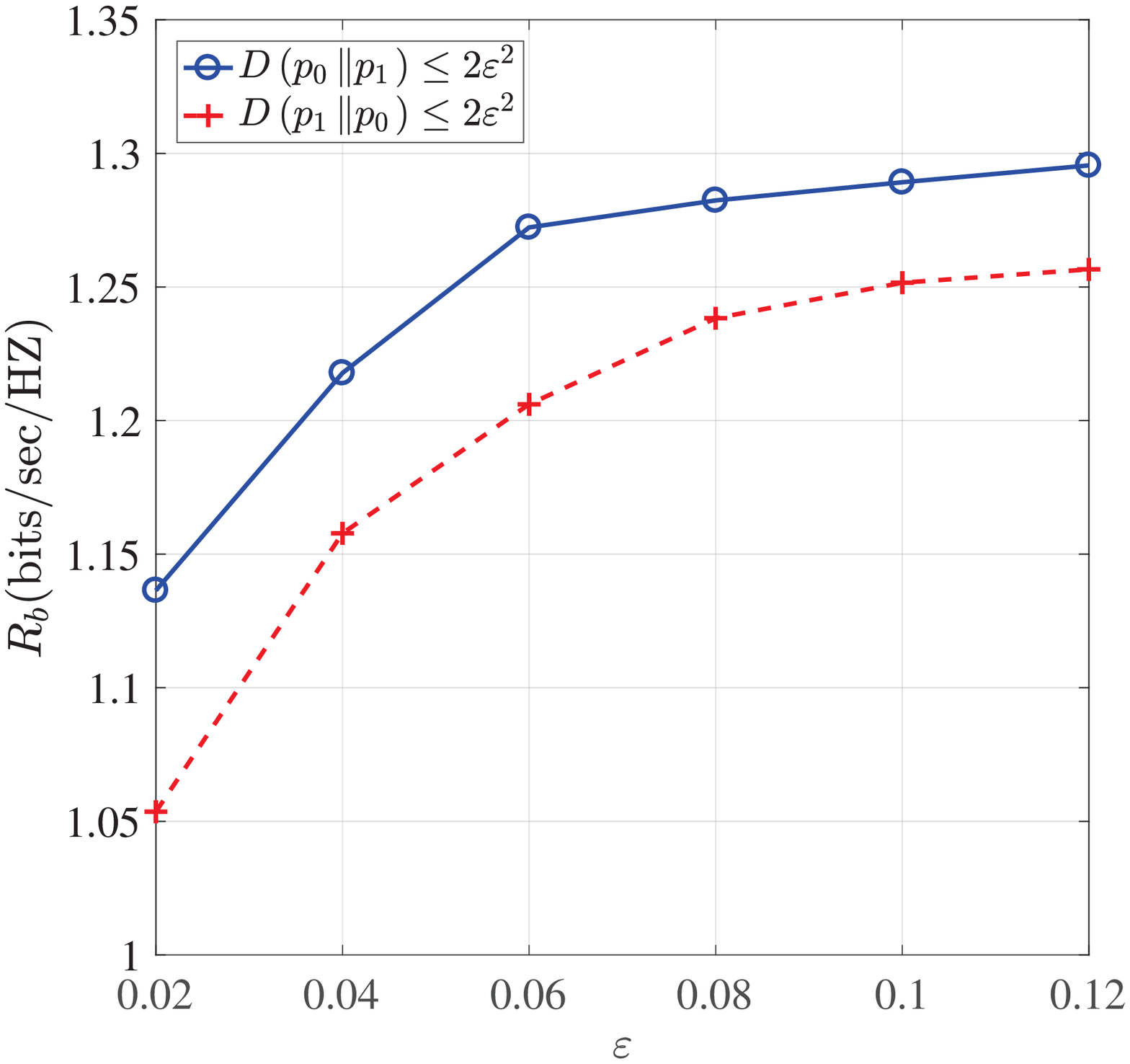}
      \vskip-0.2cm\centering {\footnotesize (a)}
    \end{minipage}
     \begin{minipage}[b]{0.45\textwidth}
      \centering
      \includegraphics[height=7cm,width=8cm]{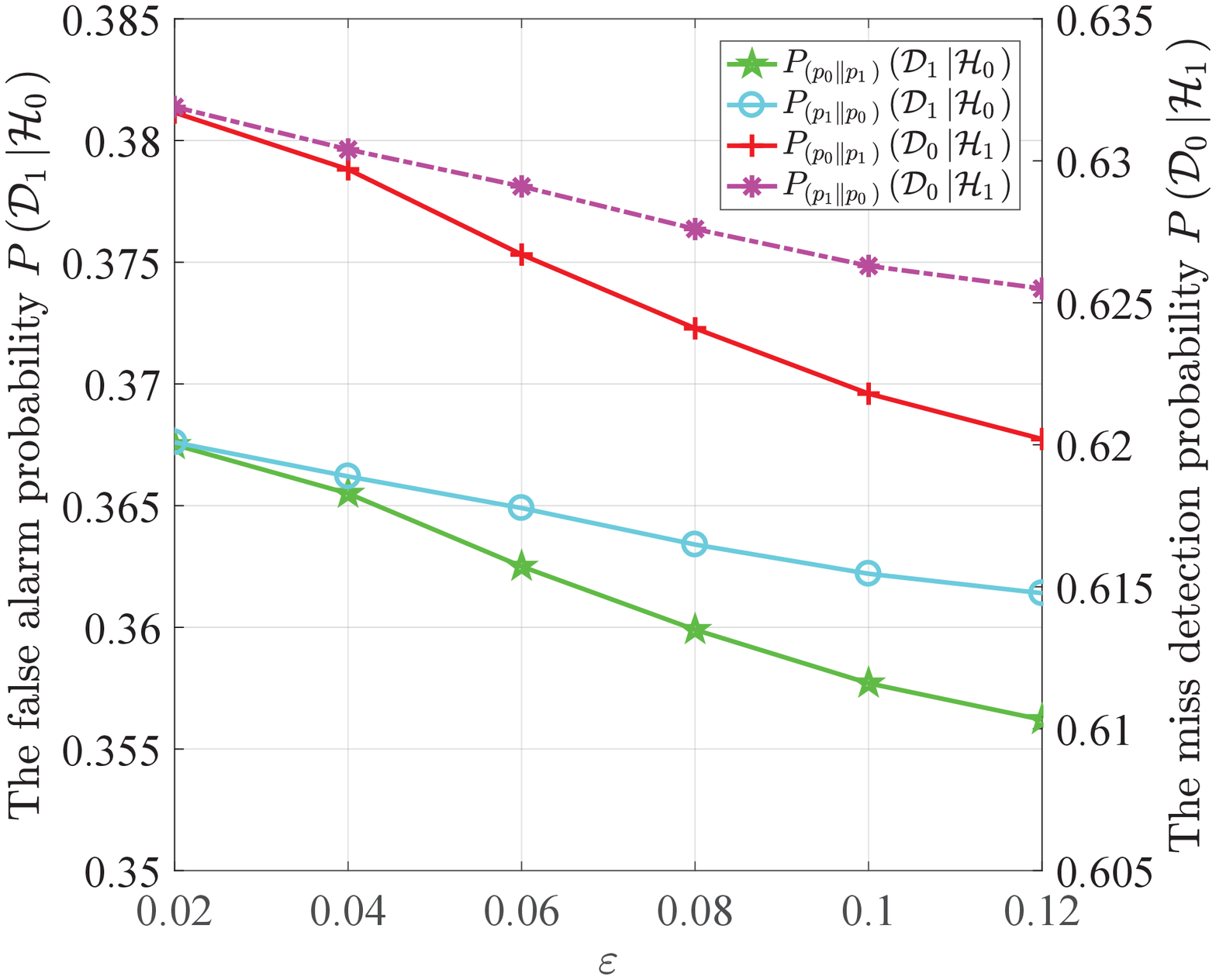}
      \vskip-0.2cm\centering {\footnotesize (b)}
    \end{minipage}\hfill
 \caption{The value of $\varepsilon$ versus (a) the covert rate and (b) the detection error probabilities  with CSI errors   $v_w=0.005$.}
 \label{fig5}  
\end{figure}

\begin{figure}
    \begin{minipage}[b]{0.45\textwidth}
      \centering
      \includegraphics[height=7.5cm,width=7.5cm]{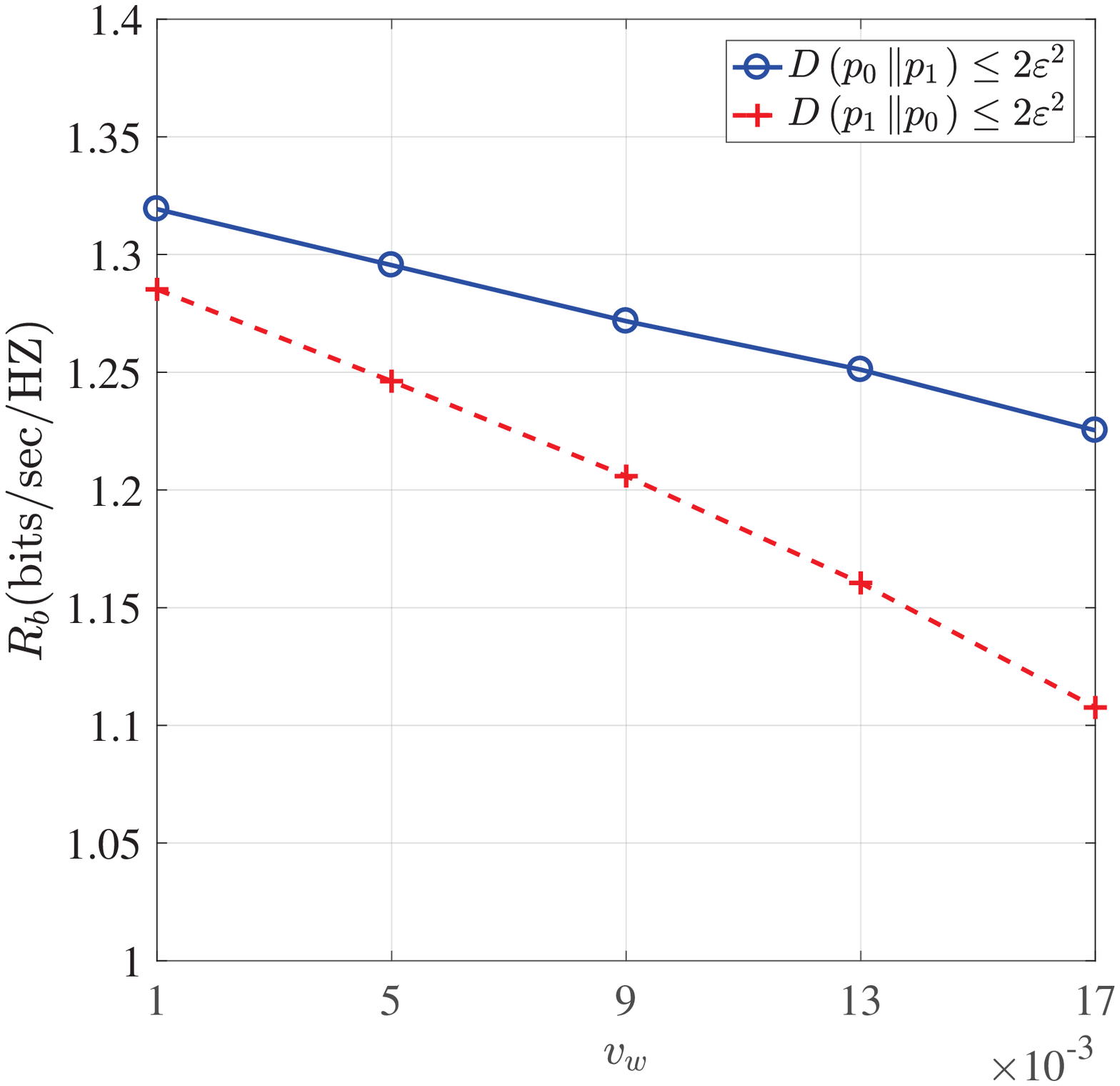}
      \vskip-0.2cm\centering {\footnotesize (a)}
    \end{minipage}
     \begin{minipage}[b]{0.45\textwidth}
      \centering
      \includegraphics[height=7.5cm,width=8.5cm]{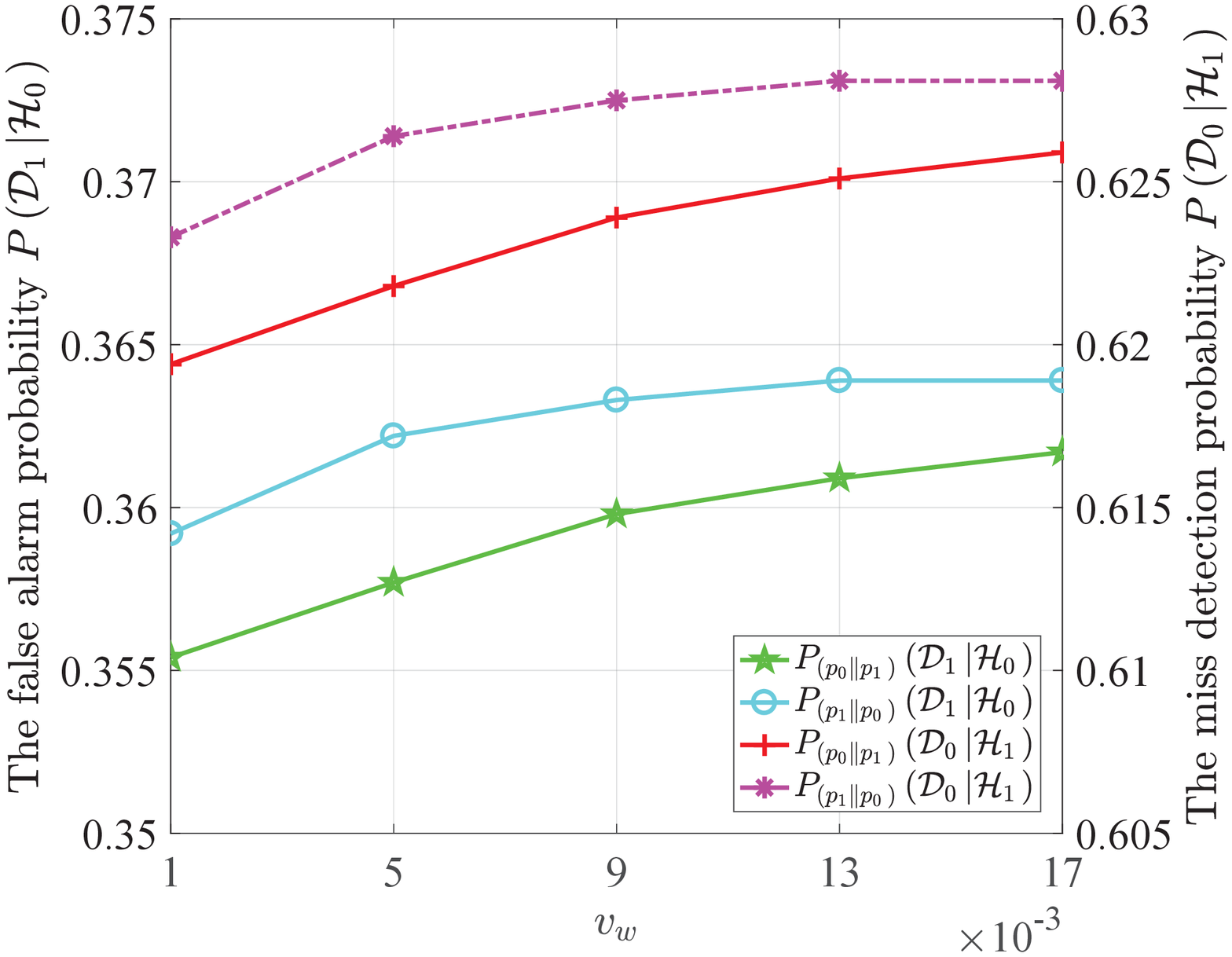}
      \vskip-0.2cm\centering {\footnotesize (b)}
    \end{minipage}\hfill
 \caption{(a) The covert rate and (b) the detection error probabilities
versus CSI errors $v_w$ with the value of $\varepsilon=0.1$.
  }
 \label{fig6}  
\end{figure}

 Fig. \ref{fig5} (a) plots   covert rates $R_{\rm{b}}$ versus the value of $\varepsilon$ for the two KL divergence cases
 with CSI errors   $v_w=0.005$, where ${P_{\left( {{p_0}\left\| {{p_1}} \right.}\right)}} \left( {{{\cal D}_1}\left| {{{\cal H}_0}} \right.} \right)$ represents the false alarm probability $P \left( {{{\cal D}_1}\left| {{{\cal H}_0}} \right.} \right)$ in the case of $D\left( {{p_0}\left\| {{p_1}} \right.} \right) \le 2{\varepsilon ^2}$, and the other notation is defined likewise.
 Such simulation result is consistent with the theoretical analysis showing that when $\varepsilon$ becomes larger, the covertness constraint is more loose, which causes $R_{\rm{b}}$ to become larger. And the rate of performance improvement also decreases in Fig. \ref{fig5} (a) with increasing $\varepsilon$.
   Fig. \ref{fig5} (b) plots the false alarm probability $P \left( {{{\cal D}_1}\left| {{{\cal H}_0}} \right.} \right)$ and the missed detection probability $P \left( {{{\cal D}_0}\left| {{{\cal H}_1}} \right.} \right)$ versus the value of $\varepsilon$ with CSI errors   $v_w=0.005$.
 We observe that  under either case of the covertness constraint, the false alarm probability $P \left( {{{\cal D}_1}\left| {{{\cal H}_0}} \right.} \right)$ and the missed detection probability $P \left( {{{\cal D}_0}\left| {{{\cal H}_1}} \right.} \right)$ are decreasing as    $\varepsilon$ increases, where $P \left( {{{\cal D}_1}\left| {{{\cal H}_0}} \right.} \right)$ is always lower than $P \left( {{{\cal D}_0}\left| {{{\cal H}_1}} \right.} \right)$.
It implies that when the convert constraint is looser, the detection performance of Willie becomes better.
  Moreover,   Fig. \ref{fig5} (b) also verifies    the effectiveness of the proposed robust beamformers design in covert communications, i.e.,
 $ \Pr \left( {{{\cal D}_1}\left| {{{\cal H}_0}} \right.} \right) + \Pr \left( {{{\cal D}_0}\left| {{{\cal H}_1}} \right.} \right) \ge 1 - \varepsilon$.
 Therefore, from Fig. \ref{fig5}, we reveal the tradeoff between Willie's detection performance and Bob's covert rate, and a desired tradeoff can be achieved via a proper robust beamformer design.

\begin{figure}[h]
      \centering
	\includegraphics[width=7.5cm]{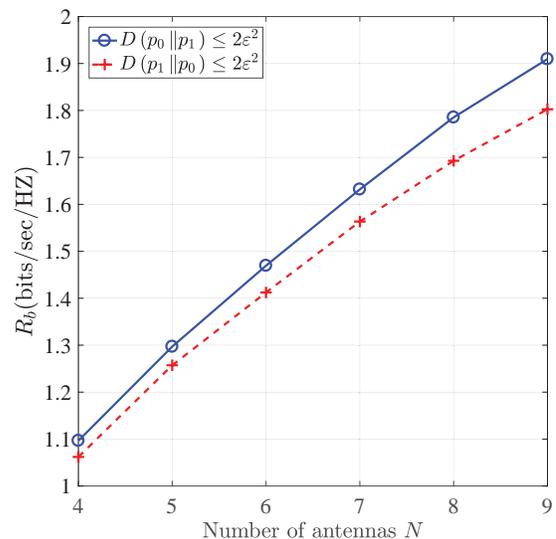}
 \caption{ Covert rates $R_{\rm{b}}$ versus  number of  antennas $N$  with CSI errors   $v_w=0.005$.}
  \label{2_N_rb} 
\end{figure}

 Fig. \ref{fig6} (a) plots   covert rates $R_{\rm{b}}$ versus CSI errors $v_w$ under
 two covertness constraints $D\left( {{p_0}\left\| {{p_1}} \right.} \right) \le 2{\varepsilon ^2}$ and $D\left( {{p_1}\left\| {{p_0}} \right.} \right) \le 2{\varepsilon ^2}$.
 We observe that as  $v_w$ increases, the  covert rates $R_{\rm{b}}$ of  two covertness constraints decrease, and the rates gap increases.
    Fig. \ref{fig6} (b) plots the false alarm probability $P \left( {{{\cal D}_1}\left| {{{\cal H}_0}} \right.} \right)$ and the missed detection probability $P \left( {{{\cal D}_0}\left| {{{\cal H}_1}} \right.} \right)$ versus  CSI errors $v_w$ under
 two covertness constraints $D\left( {{p_0}\left\| {{p_1}} \right.} \right) \le 2{\varepsilon ^2}$ and $D\left( {{p_1}\left\| {{p_0}} \right.} \right) \le 2{\varepsilon ^2}$.
 We observe that  under the two cases of covertness constraint  
 both the false alarm probability $P \left( {{{\cal D}_1}\left| {{{\cal H}_0}} \right.} \right)$ and the missed detection probability $P \left( {{{\cal D}_0}\left| {{{\cal H}_1}} \right.} \right)$ increase as    $v_w$ increases, where $P \left( {{{\cal D}_1}\left| {{{\cal H}_0}} \right.} \right)$ is always lower than $P \left( {{{\cal D}_0}\left| {{{\cal H}_1}} \right.} \right)$.
 Moreover, Fig. \ref{fig6} implies that a large error $v_w$  may lead to a bad beamformer design in terms  of cover rate $R_{\rm{b}}$. However, such beamformer may confuse the detection of Willie, which is also good for Bob. Therefore, such tradeoff also should be paid attention in the beamformer design.
Moreover, the detection error probabilities do not continuously increase with increasing $v_w$.

  Finally, Fig. \ref{2_N_rb} shows  the covert rates $R_{\rm{b}}$ versus the number of  antennas $N$ for
  two covertness constraints $D\left( {{p_0}\left\| {{p_1}} \right.} \right) \le 2{\varepsilon ^2}$ and $D\left( {{p_1}\left\| {{p_0}} \right.} \right) \le 2{\varepsilon ^2}$, where ${{{{\left\| {{{\bf{w}}_{c,0}}} \right\|}^2}}}=1\rm{dBW}$, $\varepsilon=0.1$ and $v_w=0.005$.
From  Fig. \ref{2_N_rb}, we can see that the higher the number of antennas $N$ is, the higher the achieved covert rates $R_{\rm{b}}$ will be, which is similar to the case in Fig. \ref{N_rb}.
  From Fig. \ref{fig5}-\ref{2_N_rb}, we observe  the rates with the covertness constraint $D\left( {{p_0}\left\| {{p_1}} \right.} \right) \le 2{\varepsilon ^2}$
 are higher than those with   the covertness constraint  $D\left( {{p_1}\left\| {{p_0}} \right.} \right) \le 2{\varepsilon ^2}$. This is because
    $D\left( {{p_1}\left\| {{p_0}} \right.} \right) \le 2{\varepsilon ^2}$ is stricter than
   $D\left( {{p_0}\left\| {{p_1}} \right.} \right) \le 2{\varepsilon ^2}$, and  this conclusion is also verified in \cite{Yan2019Gaussian}.

\section{Conclusions}

In this paper, we designed   a   covert beamformer, ZF beamformer and robust
beamformer  for  covert communication  networks, where  the communication link with
Carol is exploited  as a cover.
  For the   perfect WCSI scenario, we
develop both the  covert beamformer design and low-complexity ZF beamformer design to   maximize the covert rate.
Furthermore,  to quantify
the impact of practical channel estimation errors,  we considered the   imperfect WCSI scenario, and  proposed   robust beamformers design,
which can maximize the covert rate   while meeting covert requirements.
In addition,
 to evaluate the performance of the robust beamformers design, we derived
 the covert decision threshold of Willie, and false alarm
probability and missed detection probability expressions.
Numerical results illustrated the validity of the proposed  beamformers design and
provide useful insights on the impact of the involved system design parameters
on the covert communications performance.
 In the future,  we would  further   investigate the covert communications where  Willie is equipped with multi-antenna.

  \begin{appendices}

\section{Proof of  Lemma $1$}

 \emph{Proof.}
We rewrite function $g(r_b)$ as following compact form:
\begin{subequations}
\begin{align}
f\left( x \right) =\mathop{\max }\limits_{\bf{W}\in\mathcal{W}}&~ x
\\
{\rm{s}}.{\rm{t}}. \quad & a(\mathbf{W})\ge x b(\mathbf{W}),
\end{align}
\end{subequations}
where $\bf{W}:=[{{\bf{W}}_{\rm{b}}},{{\bf{W}}_{{\rm{c}},1}}]$, $a(\bf{W}):=\phi({\bf{W}}_{\rm{b}})$, $b(\bf{W}):=\theta({\bf{W}}_{{\rm{c}},1})$, $x \ge 0$.

Next, we will examine the concavity of function $f(x)$ over $x\ge0$ by definition below. First, for $0 \le \theta  \le 1$ and $x_1,x_2\ge 0$, we have
\begin{subequations}\label{A_C3}
\begin{align}
&f\left( {\theta {x_1} + \left( {1 - \theta } \right){x_2}} \right)
\\
=&\mathop {\max }\limits_{\mathbf{W}\in\mathcal{W}}~ \theta {x_1} + \left( {1 - \theta } \right){x_2}
\\
&{\rm{s}}.{\rm{t}}.~a\left( {\mathbf{W}} \right)  \ge \left( {\theta {x_1} + \left( {1 - \theta } \right){x_2}} \right)b\left( {\mathbf{W}} \right),\label{A_C3b}
\end{align}
\end{subequations}

Then,  we have functions $\theta f\left( {{x_1}} \right)$  and $\left( {1 - \theta } \right)f\left( {{x_2}} \right)$  as follows
\begin{subequations}\label{A_C1}
\begin{align}
\theta f\left( {{x_1}} \right) =\mathop {\max }\limits_{\bf{W}\in\mathcal{W}} &~\theta {x_1}\\
{\rm{s}}.{\rm{t}}.~&a\left( {\mathbf{W}} \right)  \ge {x_1}b\left( {\mathbf{W}} \right),
\end{align}
\end{subequations}
\begin{subequations}\label{A_C2}
\begin{align}
\left( {1 - \theta } \right)f\left( {{x_2}} \right) =\mathop {\max }\limits_{\bf{W}\in\mathcal{W}}~& \left( {1 - \theta } \right){x_2}\\
{\rm{s}}.{\rm{t}}.&~a\left( {\bf{W}} \right) \ge {x_2}b\left( {\bf{W}} \right).
\end{align}
\end{subequations}

Let $c\left( {\bf{W}} \right) \buildrel \Delta \over = \frac{{a\left( {\bf{W}} \right)}}{{b\left( {\bf{W}} \right)}}$. We   have
\begin{subequations}\label{A_C4}
\begin{align}
\theta f\left( {{x_1}} \right) + \left( {1 - \theta } \right)f\left( {{x_2}} \right) =\mathop {\max }\limits_{\bf{W}\in\mathcal{W}}& ~\theta {x_1} + \left( {1 - \theta } \right){x_2}\\
{\rm{s}}.{\rm{t}}.&0 \le {x_1} \le c\left( {\bf{W}} \right),\\
&0 \le {x_2} \le c\left( {\bf{W}} \right).
\end{align}
\end{subequations}

Note that constraint \eqref{A_C3b} can be equivalently written as
\begin{align} \theta {x_1} + \left( {1 - \theta } \right){x_2} &\le c\left( {\bf{W}} \right),
\end{align}
where $x_1,x_2\ge 0$.

It can be easily checked that when $0\le\theta\le 1$, the feasible region of $x_1$ and $x_2$  shown in \eqref{A_C3b} is larger than that in  \eqref{A_C4}. Therfore, we have
\begin{equation}\label{A_C7}
\theta f\left( {{x_1}} \right) + \left( {1 - \theta } \right)f\left( {{x_2}} \right) \le f\left( {\theta {x_1} + \left( {1 - \theta } \right){x_2}} \right),
\end{equation}
implying that $f\left( x \right)$   is concave in $x$.
In other words, function \eqref{C_15} is concave in ${r_{\rm{b}}}$. \hfill $\blacksquare$
\end{appendices}

\bibliographystyle{IEEE-unsorted}
\bibliography{Robust_covert}

\end{document}